\documentclass[sigconf]{acmart}

\usepackage{footmisc}

\usepackage[inline]{enumitem}
\usepackage{float}
\usepackage{listings,multicol}

\AtBeginDocument{\DeclareCaptionSubType{lstlisting}}

\usepackage{newfloat,caption}
\usepackage{subcaption}

\setcopyright{none}
\renewcommand\footnotetextcopyrightpermission[1]{} 

\newfloat{lstfloat}{htbp}{lop}
\floatname{lstfloat}{Listing}

\usepackage[capitalise,noabbrev]{cleveref}
\crefname{lstlisting}{listing}{Listings}
\Crefname{lstlisting}{Listing}{Listings}
\Crefname{sublstlisting}{Listing}{Listings}
\Crefname{sublstlisting}{Listing}{Listings}

\usepackage{siunitx}

\usepackage{xcolor}
\definecolor{sd_darkblue}{HTML}{05495f}
\definecolor{sd_mediumblue}{HTML}{007fa4}
\definecolor{sd_orange}{HTML}{ef6a3f}
\definecolor{sd_green}{HTML}{cadb2e}
\definecolor{sd_cyan}{HTML}{9ECDC3}

\AtBeginDocument{%
  }

\begin{document}

\title{BARQ: A Vectorized SPARQL Query Execution Engine}

\author{Simon Grätzer}
\affiliation{%
  \institution{Stardog Union}
  \city{New York}
  \country{USA}
}
\email{simon.graetzer@stardog.com}

\author{Lars Heling}
\orcid{0000-0001-9668-8935}
\affiliation{%
  \institution{Stardog Union}
  \city{New York}
  \country{USA}
}
\email{lars.heling@stardog.com}

\author{Pavel Klinov}
\affiliation{%
  \institution{Stardog Union}
  \city{New York}
  \country{USA}
}
\authornote{The corresponding author}
\email{pavel@stardog.com}

\renewcommand{\shortauthors}{Grätzer et al.}

\begin{abstract}
Stardog is a commercial Knowledge Graph platform built on top of an RDF graph database whose primary means of communication is a standardized graph query language called SPARQL. 
This paper describes our journey of developing a more performant query execution layer and plugging it into Stardog's query engine. 
The new executor, called BARQ, is based on the known principle of processing batches of tuples at a time in most critical query operators, particularly joins. 
In addition to presenting BARQ, the paper describes the challenges of integrating it into a mature, tightly integrated system based on the classical tuple-at-a-time Volcano model. 
It offers a gradual approach to overcoming the challenges that small- to medium-size engineering teams typically face. 
Finally, the paper presents experimental results showing that BARQ makes Stardog substantially faster on CPU-bound queries without sacrificing performance on disk-bound and OLTP-style queries.
\end{abstract}


\maketitle
\pagestyle{plain}

\section{Introduction}
\label{sec:introduction}
Originally released in 2011, Stardog is a Knowledge Graph platform built around a database system for the graph data model called RDF and the (sub)graph matching query language called SPARQL (c.f. \cref{sec:rdf}). 
Over the years Stardog has evolved from a relatively simple RDF/SPARQL database into a large system that offers a wide range of functionality in the data integration space, namely, data virtualization, deductive reasoning, and machine learning. 
Recently, it also started serving as the backend for a generative AI data assistant called Voicebox\footnote{\url{https://stardog.ai/}}.

Nonetheless, all functionality ultimately relies on Stardog's ability to execute database queries in a fast and scalable manner. 
The storage and query engine are at the heart of the system and they must evolve to support workloads from all those different components. 
The workloads have been changing in both intensity and nature over the years, hence the database inside Stardog must adapt, too. 
The storage engine went through a major overhaul in 2020 migrating from a read-optimized, home-grown B+ tree-like system to a more read-write balanced layer based on RocksDB \cite{DBLP:conf/cidr/DongCGBSS17}. Now the time has come to modernize the execution engine, which, from the inception point, has been based on the following principles:

\begin{enumerate}
\item Pull-based, bottom-up evaluation (aka the "Volcano model" \cite{DBLP:journals/tkde/Graefe94}). Each operator in the execution plan pulls data from its argument or child operators.
\item Tuple-at-a-time evaluation (aka row-based). Each operator processes just enough argument tuples to produce a single output tuple.
\end{enumerate}

\begin{figure}[t]
    \centering
    \includegraphics[width=0.35\textwidth]{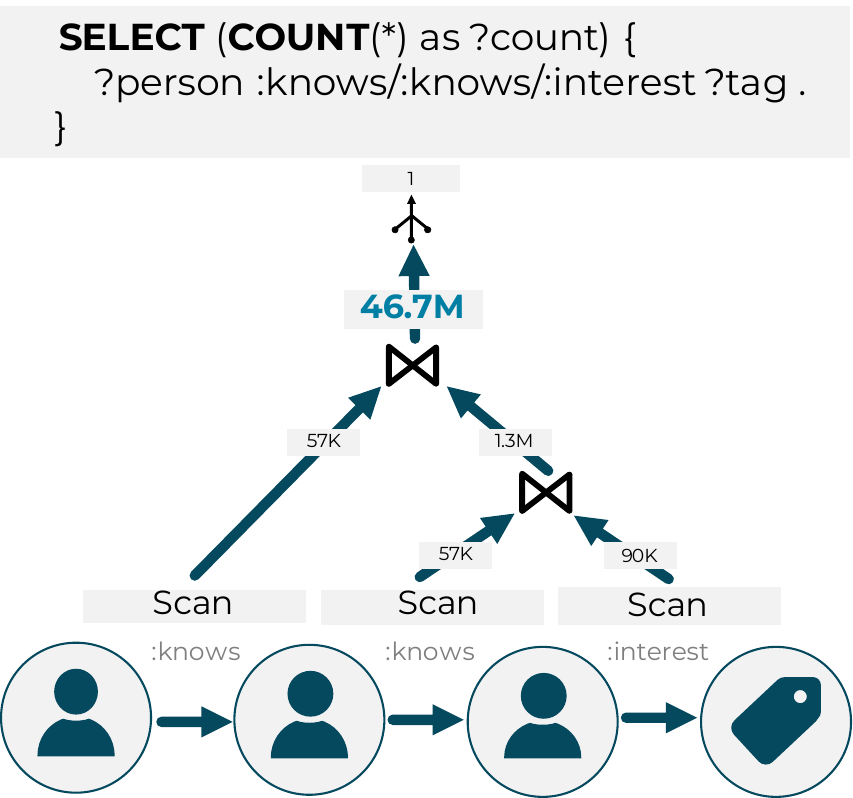}
    \caption{Motivating Example}
    \label{fig:motivation}
\end{figure}

The tuple-at-a-time model has worked well for executing specific, selective, OLTP-style queries, where the key is avoiding unnecessary disk and network IO. Stardog’s storage layer maintains various sorted indexes over the graph edges (RDF quads). Assuming the optimal query plan, the traditional model supports skips over data ranges on disk that do not satisfy join conditions in the query. Plan optimality is, of course, a crucial assumption here, that's why Stardog has invested heavily in a cost-based query optimizer. Specifically, the join order optimizer relies on a wide range of cardinality estimation techniques, including graph summarization, count-min sketches, and statistics collection from remote data sources in case of virtualization. However, the row-based model is not the best approach to dealing with CPU-bound queries, as the following example shows.

\subsubsection*{Motivating Example}
Consider a simple SPARQL query shown in \cref{fig:motivation}. It’s a simplified version of query 6 from the LSQB benchmark that selects all pairs of people in a social network dataset connected via directed 2-hop paths, fetches interest tags for one person in each pair, and counts the results\footnote{For brevity, we use \texttt{:knows} in examples throughout the paper instead of the 
\texttt{:Person\_knows\_Person} IRI in the LSQB dataset}. 
Query planning (e.g. join order optimization) for this query is straightforward.
However, the number of 2-hop paths grows extremely fast in dense graphs. Specifically, even on a small graph with less than 10M edges (LSQB scale factor of 0.3), the number of tuples to be counted at the top of the plan is more than 46.7M. The explosion happens in joins, all intermediate tuples are needed, and thus optimizing disk IO will not improve performance. 

Consequently, the time has come to address this problem rather than compensate by improving other parts of the system. 
In the remainder of the paper, we present the steps we have taken along this journey, called project BARQ (\textbf{B}atch-based \textbf{A}ccele\textbf{R}ated \textbf{Q}uery Executor), particularly towards integrating the new execution layer into the existing query engine. 

\subsubsection*{Contributions}
The paper makes the following two contributions.
\begin{enumerate}
\item It presents Stardog’s new batch-based (sometimes called "vectorized") execution layer and evaluates its performance on various workloads in comparison to the existing query executor.
\item It describes our journey of migrating the main query execution pieces to this batch-based model. The key aspect of that process is that it should be in a pragmatic and incremental way because the bottleneck of re-implementing a key component in a large, tightly integrated, and mature system is nearly always the engineering resources. 
\end{enumerate}

Stardog has a small query engine development team and therefore we decided to focus on upgrading one specific aspect of the query engine while keeping others, like the pull-based evaluation model and the query optimizer, intact. 
We also concluded, for reasons explained in \cref{sec:integration}, that both tuple-at-a-time and vectorized execution must co-exist, at least temporarily. 
While these decisions certainly helped to keep the scope of the project under control and deliver it on time, they have their own technical challenges. 
Overcoming these challenges offers general lessons for other small-to-medium size system vendors.

The paper is structured as follows. After providing a short background on RDF/SPARQL and Stardog's architecture in \Cref{sec:background}, we delve into the two main technical sections: \Cref{sec:barq} describes the main concepts and optimizations behind BARQ while \Cref{sec:integration} goes into the details of integrating it into the rest of the query engine. \Cref{sec:evaluation} then discusses the results of our experimental evaluation. In \Cref{sec:related_work} discusses the related work. Finally, \Cref{sec:conclusion} concludes the paper.

\section{Background}
\label{sec:background}
\subsection{RDF and SPARQL}
\label{sec:rdf}

The Resource Description Framework (RDF) is a graph-based data model standardized by the World Wide Web Consortium (W3C)\footnote{\url{https://www.w3.org/TR/rdf11-concepts/}}. An RDF graph consists of a set of triples that are 3-tuples of RDF terms: $t = (s, p, o)$ with $s$ the subject, $p$ the predicate, and $o$ the object of the triple. Essentially, a triple represents a directed, labeled edge in the graph. RDF 1.1 supports the notion of \emph{dataset} allowing users to place their triples in different named graphs (also called \textit{context}). A triple in a named graph is sometimes called a \emph{quad}. RDF is a schema-less data model which does not enforce a strict type system. 
A simple RDF graph $G$ is \texttt{\{(:Alice, :knows, :Bob ), (:Alice, :knows, :Charlie), (:Bob, :worksAt, :ACME )\}}.

SPARQL is the standard query language for RDF and is supported by all major RDF databases (also called triple/quad stores)\footnote{\url{https://www.w3.org/TR/sparql11-query/}}. The SPARQL query language allows for specifying  template-based graph queries which are evaluated using a pattern matching approach. The atomic elements in SPARQL are triple patterns. Triple patterns are 3-tuples that consist of RDF terms and variables. For example,  \texttt{(:Alice, :knows, ?person)}  is a triple pattern to retrieve everyone `:Alice` knows. The evaluation of a triple pattern requires finding all mappings (so-called solution mappings) from variables to RDF terms such that replacing the variables with the RDF terms yields a triple in the graph. For example, evaluating the example triple pattern over $G$ yields the following solution mappings: \texttt{\{\{?person $\to$ :Bob\}, \{?person $\to$ :Charlie\}\}}. In the remainder of this work, we use the terms tuples or rows synonymously with solutions.

Using triple patterns and a set of algebraic operators as building blocks, more complex graph patterns such as joins, unions, left-outer joins can be constructed. The most common graph patterns are Basic Graph Patterns (BGPs) which are sets of triple patterns. For example, we can find every person Alice knows as well as the company they work for with the following BGP $B$ =\texttt{\{(:Alice :knows ?person), (?person :worksAt ?company)\}}.

In principle, any algorithm that determines all homomorphisms from the BGP (query graph) to the RDF graph can be used to retrieve the solution mappings for the BGP.  Stardog treats BGPs as first-order conjunctive queries and evaluates them using join operators over scans (which evaluate triple patterns). The evaluation of BGPs requires joining the solutions of multiple triple patterns. Hence, evaluating $B$ over $G$ requires joining the solution mappings for each triple pattern in the BGP. For example, joining the following solution:  \texttt{\{\{ ?person $\to$ :Bob \}, \{ ?person $\to$ :Charlie \}\} $\bowtie$ \{\{ ?person $\to$ :Bob, ?company $\to$ :ACME \}\}}  yields the solution mapping of \texttt{\{\{?person $\to$ :Bob, ?company $\to$ :ACME \}\}}. Note that the joins are natural inner equi-joins i.e. all join conditions are equality checks over values of shared variables.

SPARQL also supports many additional operators such as projections, filters, aggregations, sorting, and so on to create complex queries. Their semantics as defined in the specification is relational and usually very close to SQL. For example, the \texttt{OPTIONAL} operator in SPARQL behaves similarly to left-outer joins in SQL while \texttt{MINUS} is an anti-join. That makes various query optimization and execution techniques developed for relational databases easily adaptable to SPARQL.

\subsection{Stardog Architecture}
\label{sec:architecture}

\begin{figure}[t]
    \centering
    \includegraphics[width=0.48\textwidth]{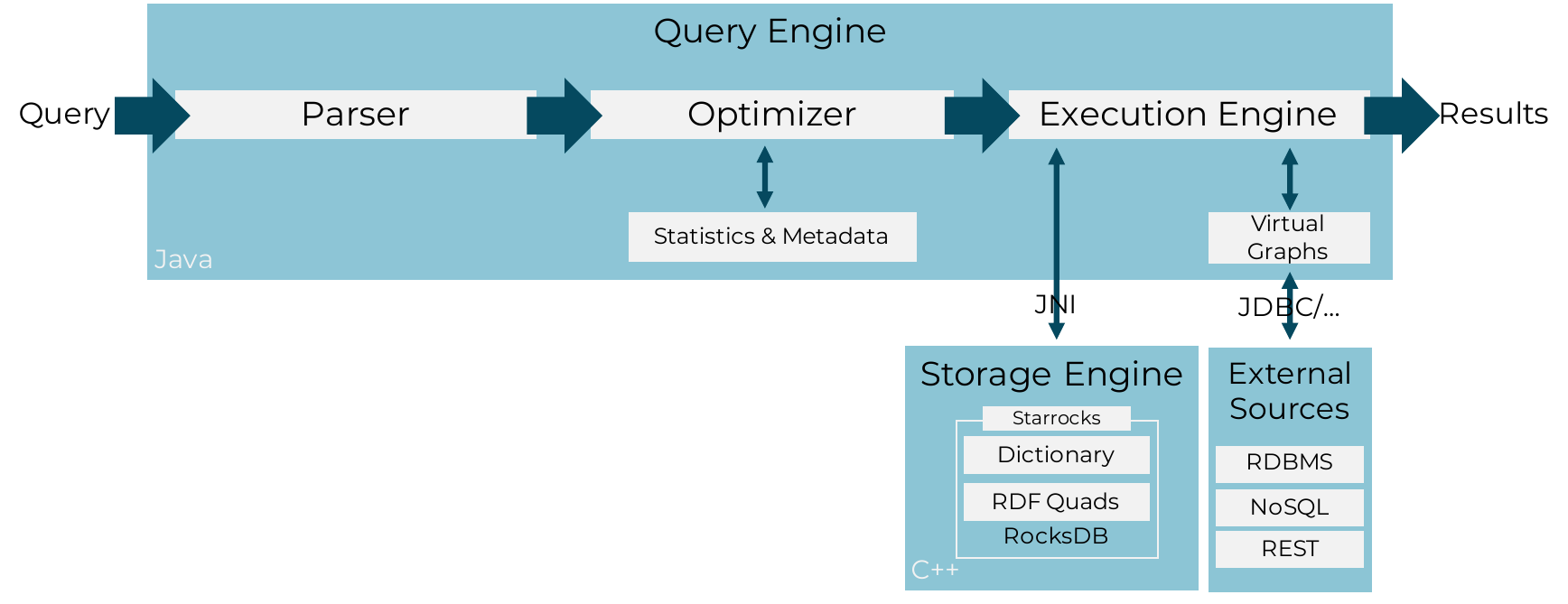}
    \caption{Stardog Architecture Overview}
    \label{fig:architecture}
\end{figure}

As shown in \cref{fig:architecture}, Stardog uses a query processing pipeline that is typical for most SQL database systems. It processes each query in consecutive stages: 
\begin{enumerate*}[label=(\arabic*)]
   \item parsing and dictionary encoding, 
   \item logical query optimization, 
   \item translation, 
   \item execution, and 
   \item result decoding. 
\end{enumerate*}
One notable difference from most SQL databases is that there is no separate binding phase that normally resolves (binds) expressions to table and column names, and obtains information about their datatypes (example: DuckDB \cite{DBLP:conf/vldb/Raasveldt22}). That part is mainly bypassed due to the schema-less nature of RDF and SPARQL (e.g., the query engine cannot assume that all values in the range of a particular predicate have a certain datatype). 

Stardog is a hybrid system where the storage engine is implemented in C++ and the rest of the system (including the query engine and ACID transaction support) in Java. Data exchange happens in batches over the JNI interface using shared memory.

\subsubsection{Storage Engine}
\label{sec:storage}

Stardog's storage engine consists of two parts: sorted indexes and the mapping dictionary. Both ideas date back to early RDF systems, such as RDF-3x\cite{DBLP:journals/pvldb/NeumannW08}. The indexes store RDF quads (graph edges) as lexicographically sorted collections of four 64-bit numbers representing the subject, predicate, object, and context of each quad. The dictionary is a key-value store bi-directionally mapping each number that occurs in some quad to the corresponding RDF term (an IRI, a literal, or a blank node). The dictionary plays a critical role: First, it makes the indexes compact, efficiently compressible, and relatively cheap to sort. It mitigates the index redundancy issue, as Stardog stores the same quads multiple times according to different order\footnote{Stardog does not sort the data according to all permutations of subject/predicate/object/context, as early RDF systems.}. Second, it ensures that most performance-critical computation during query execution, such as, joins, hashing, sorting, etc., can be done over numbers and not arbitrary datatypes (notable exceptions are \texttt{FILTER}, \texttt{BIND}, and \texttt{ORDER BY}  operators that evaluate SPARQL expressions over RDF terms). Many queries can be processed entirely over numbers with dictionary decoding happening only for final results.

Both components use RocksDB \cite{DBLP:conf/cidr/DongCGBSS17} to store data in multiple column families (akin to tables). RocksDB provides low-level primitives for data access, such as point look-ups, range scans, key/value updates, and file ingestion. The custom \texttt{starrocks} layer implemented in C++ translates between these and higher-level operations, such as SPARQL triple scans.

In addition, Stardog provides data virtualization capability allowing users to execute SPARQL queries over non-RDF data sources using mappings. It supports many SQL and non-SQL data sources, such as document databases or text search engines. The query engine accesses those using various implementations of \texttt{SERVICE} operators, for example, using JDBC connectors for relational data.

\subsubsection{Query Optimizer}
\label{sec:optimizer}  
Historically, Stardog has prioritized the improvement of the optimizer over the executor. When the typical query workload consists mostly of selective queries with specific criteria, the most important thing is to be able to select the execution plan that gets to the required data as fast as possible without loading unnecessary data from either disk or network (in case the query requests data from external endpoints, like Virtual Graphs\footnote{https://docs.stardog.com/virtual-graphs/}). In other words, ensuring that the amount of intermediate query results is small used to be more important than the throughput of intermediate operators, such as joins, that process the results. Consequently, Stardog incorporates an extensive set of optimizations, such as (but not limited to):

\begin{itemize}
    \item cost-based join reordering,
    \item algebraic transformations: pushing filters and projections, distributing joins over unions, pushing selective patterns into nested scopes, extracting repeated patterns, etc.,
    \item de-correlation of \texttt{EXISTS} and \texttt{NOT EXISTS} expressions in filters\footnote{\texttt{EXISTS} expressions in SPARQL are defined in terms of variable substitution and are similar to correlated subqueries in SQL. In many cases, their evaluation can be de-correlated by rewriting into semi-/anti-joins so that the \texttt{EXISTS} pattern is evaluated once instead of for each argument row.},
    \item static evaluation of constant expressions, constant propagation, and inlining.
\end{itemize}

In addition, the optimizer performs various transformations that are often presented as physical optimizations in other systems, for example, selecting specific join algorithms and assigning sort orders to operators.
The most critical optimizations, such as join reordering and pushing filters, are based on the cost model that takes cardinality estimations for each SPARQL pattern as input. Thus, the effectiveness of the optimizer greatly depends on the estimation accuracy. Stardog implements a wide range of techniques to estimate how many results a particular SPARQL pattern would match in the data. Most of them are based on precomputed graph statistics, such as predicate cardinality, frequent chain cardinality, characteristic sets \cite{DBLP:conf/icde/NeumannM11} enhanced with count-min sketches \cite{DBLP:journals/jal/CormodeM05}, as well as commonly used assumptions (independence) and heuristics. A more detailed overview of Stardog’s cardinality estimation framework is beyond the scope of this paper.

\subsubsection{Execution Engine}
\label{sec:executor_legacy}

The legacy execution engine follows the classic Volcano model \cite{DBLP:journals/tkde/Graefe94}. Once the optimized logical plan is produced, it is translated into a tree of \texttt{Operator} objects, each of which represents an iteration over sets of \texttt{variable $\to$ value} tuples (aka \textit{solutions} in SPARQL). Again, each \texttt{value} is a 64-bit dictionary-encoded number. The executor simply calls the \texttt{next()} method of the root \texttt{Operator}, which then (recursively) calls the same method on its child operators, all the way to the bottom. At the bottom level, there are operators that either read data from storage (index scans) or fetch it from external endpoints over network protocols. Until Stardog version 10, each operator returns a single tuple as the result of a \texttt{next()} call even if it used some data buffering internally for whatever reason, such as sorting or building a hash table (those are called \textit{pipeline breakers}).

Each operator in Stardog declares whether it produces data according to some sort order, such as sorted by values of a particular variable. If yes, it supports a second method in addition to \texttt{next()}, called \texttt{skip()}. The method accepts a single parameter — the target value of the sort key variable — and re-positions the iteration to the tuple with the same value of the sort key (or the next greater one). All scan operators in Stardog read data from sorted indexes and thus support a \texttt{skip()} method. Consequently, join algorithms that require sorted inputs, such as the merge join, are far more prevalent than in other systems (since often no separate sort operation is needed). Results of most triple pattern scans can be directly merge-joined, and many simple queries containing only SPARQL BGPs can be executed using just merge joins. That makes merge join performance critical in Stardog.

\subsubsection{Memory Management}
\label{sec:mm}

An important part of the query engine is the memory management layer for dealing with intermediate query results. It provides utilities like hash tables and sorted arrays. That decouples data structures from query operators enabling, for example, hash joins and \texttt{GROUP BY} algorithms share the same hashing mechanisms. The framework also takes care of pre-allocating memory pools, tracking memory consumption, enforcing per-query limits, keeping memory blocks away from Java's heap to avoid GC pauses, and spilling intermediate results to disk, when necessary. The memory management layer presents challenges to the query engine vectorization project since all supported data structures were designed for row-based data layout and processing.

\medskip

Over the years, we have noticed the high per-tuple overhead of most execution algorithms in Stardog (joins, aggregation, filtering, etc.) but the root performance problem was usually the sub-optimal execution plan. That is, the query was processing more intermediate tuples than necessary. Reasons for that could be many but more often than not it was inaccurate cardinality estimations yielding a bad join order. Our response has usually been to figure out where the optimizer has made a wrong choice and address that, for example, by improving the estimation framework, the cost functions, or any particular plan transformations. That, of course, does not make the per-tuple overhead go away but it offsets it by reducing the number of processed tuples.

\section{The BARQ Execution Engine}
\label{sec:barq}
The main goal of BARQ is improving the execution engine throughput on CPU- and memory-bound query workloads by reducing the overhead of per-tuple processing. \Cref{lst:motivation} shows the Stardog profiler\footnote{Stardog has an instrumentation-based profiler that collects runtime statistics about each operator in the query plan: \url{https://docs.stardog.com/operating-stardog/database-administration/managing-query-performance\#sparql-profiler}\label{fn:profiler}} output for the query from our motivating example in \cref{fig:motivation} to demonstrate the overhead of the legacy execution engine.

\begin{lstfloat}[t]
\begin{lstlisting}[
%aboveskip=\smallskipamount,
%belowskip=\smallskipamount,
basicstyle=\ttfamily\tiny,
%language=SPARQL, 
mathescape=true,
morekeywords={Scan,MergeJoin,Scan,Sort,Group,Filter},
commentstyle=\color{gray},
sensitive=true,
firstnumber=0, 
numbersep=5pt,
label={lst:motivation},
caption={Motivating Example: Query Profile},
numberstyle=\scriptsize,
frame=none,
escapeinside={<@}{@>},
numberfirstline=false
]
Group(aggregates=[(COUNT(*) AS ?count)]) results: 1, wall time: 16.0%
`- Filter(?person1 != ?person3) results: 46.7M, wall time: 42.3%
   `- MergeJoin(?person2) <@\textcolor{sd_mediumblue}{\textbf{results: 46.7M, wall time: 28.9\%}}@>
      +- Scan(?person1, :knows, ?person2), results: 54K, wall time: 0.1%
      `- Sort(?person2), results: 6.8M, wall time: 11.8%
         `- MergeJoin(?person3), <@\textcolor{sd_mediumblue}{\textbf{results: 1.3M, wall time: 0.6\%}}@>
            +- Scan(?person3, :interest, ?tag), results: 77K, wall time: 0.2%
            `- Scan(?person2, :knows, ?person3), results: 57K, wall time: 0.1%
\end{lstlisting}
\end{lstfloat}

The legacy merge joins process \texttt{1.3M} and \texttt{46.7M} tuples, one by one, accounting for nearly 30\% of the execution time. Note that each of those joins generates orders of magnitude more tuples than it receives from the child operators. This means that most of the time is spent (in-memory) processing and not reading from disk. Also, each output tuple requires a virtual \texttt{next()} call from the parent operator to process it further, making even trivial operators, like the inequality filter or the tuple counting in the above example, expensive. This problem cannot be addressed using alternative query plans; thus, a better performing executor is needed.

In the remainder of this section, we first discuss the main design decision and then present core concepts in BARQ that are specifically tailored to the Stardog storage layer.

\subsection{Design Decisions}
\label{sec:design_decisions}

\subsubsection*{Vectorized Execution or Code Generation}
\label{sec:vectorized_query_processing}

There are two main ways in which modern database systems execute queries: vectorized data processing and data-centric code generation \cite{DBLP:journals/pvldb/KerstenLKNPB18}. Prominent examples include MonetDB/X100 \cite{DBLP:conf/cidr/BonczZN05}, DuckDB \cite{DBLP:conf/vldb/Raasveldt22}, Photon \cite{DBLP:conf/sigmod/BehmPAACDGHJKLL22}, and Velox \cite{DBLP:journals/pvldb/PedreiraEBWSPHC22} in the former camp and Umbra \cite{DBLP:journals/pvldb/Neumann11}, and Spark SQL \cite{DBLP:conf/sigmod/ArmbrustXLHLBMK15} in the latter. Vectorized engines process data in batches to amortize the interpretation overhead of virtual function calls over the entire batch. Code generation systems avoid virtual function calls completely by using a JIT compiler to translate query plans into executable code.

We picked vectorization as the computational paradigm for BARQ because it aligns well with other components of Stardog’s query engine, including important tools such as the query profiler. Observing query execution and profiling is straightforward since the operator tree closely matches the optimized logical plan. Moreover, it has been noted that with code generation it is challenging to debug compiled query programs to find bugs or performance bottlenecks because the operator code may be fused into a single processing loop over input tuples \cite{DBLP:journals/pvldb/KerstenLKNPB18}. That would have required spending more engineering effort on debugging tools and likely delayed bringing BARQ into production.

Ultimately, both approaches have been shown to perform well on CPU-bound analytical queries \cite{DBLP:journals/pvldb/KerstenLKNPB18}, so raw performance is not the decisive factor for the choice. Also, choosing vectorization does not preclude the use of code generation in specific parts of the execution engine later. For example, often used SPARQL expressions, such as numerical range checks or string functions, can be compiled to make specific operators, such as \texttt{FILTER}, more efficient without blurring operator boundaries. This is one direction for future work.

\subsubsection*{Vectorized Operator API}
\label{sec:vectorized_operatir_api}

BARQ operators use a slightly different \texttt{Operator} API where the \texttt{next()} method returns a batch of tuples. This is more efficient than treating batches as simple arrays of the same tuples on which the legacy API is based because it allows us to use the columnar data layout (cf. \cref{sec:columnwise_solution_batches}). In particular, it is possible to evaluate the join condition or a filter expression that requires only one column without reading other columns. In addition, many common aggregation functions, such as \texttt{count}, \texttt{min}, \texttt{max}, \texttt{average}, make it possible to aggregate data within a batch and merge the results across batches.

API differences aside, BARQ query plans are executed in the same fashion. We use the "Vector Volcano" model \cite{DBLP:conf/vldb/Raasveldt22}, i.e., each operator pulls batches of tuples from the child operators using  \texttt{next()/skip()} and \texttt{reset()} calls. Unlike most other query engines, BARQ combines the use of batching with the \texttt{skip()} method because it is key to efficient processing of sorted data (not only in merge joins, but also for some forms of grouping and aggregation, for example, deduplication).

\subsubsection*{Columnar Solution Batches}
\label{sec:columnwise_solution_batches}

The columnar data layout generally works better for analytical workloads, such as aggregation. We have chosen it for BARQ mostly because it suits SPARQL joins well since those typically operate on a single (shared) variable. Conceptually, each batch is a list of rows where each row is a SPARQL solution, i.e., a set of variable-value mappings (see \cref{sec:rdf}). However, the batch stores the data in a collection of columns each corresponding to a variable in the query. A column is a \texttt{long[]} array since the RDF terms are always dictionary encoded.
This dense storage is better suited for modern CPUs with deep instruction pipelines and tiered cache architectures \cite{DBLP:journals/pvldb/LambFVTVDB12}. It enables engine designers to design processing kernels which work directly on individual single columns. Operators can be implemented as tight loops over columns of variables, enabling CPUs to use efficient data pipelining and pre-fetching. Pivoting between rows and columns can be done efficiently if needed.

BARQ implements a lightweight batch pool to reuse batches discarded during execution. Reasons for discarding batches include skipping beyond the current batch or filtering out all rows in the batch in a \texttt{FILTER} operator (see the next subsection).

\subsubsection*{Selection Vector \& Inactive Rows}
\label{sec:representing_inactive_rows}

\begin{figure}
    \centering
    \includegraphics[width=0.31\textwidth]{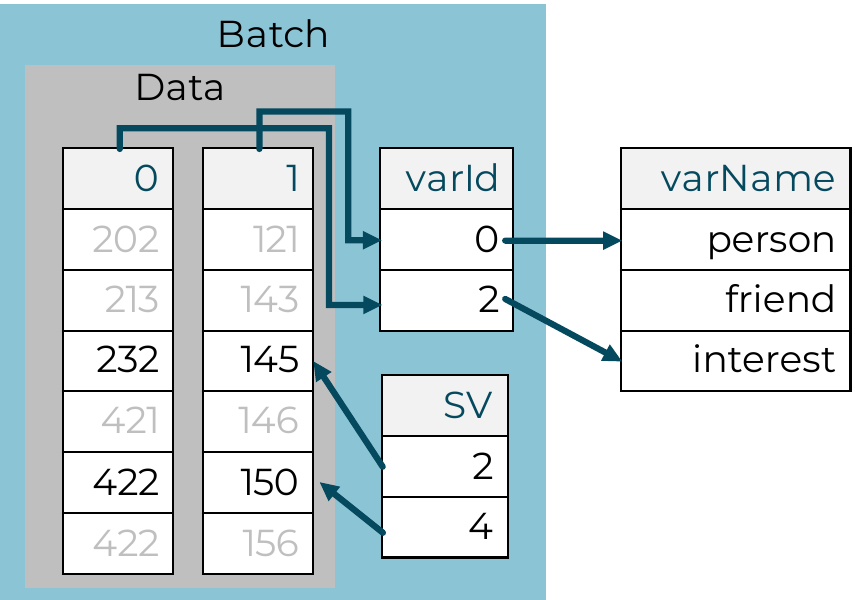}
    \caption{A column batch in BARQ}
    \label{fig:batch}
\end{figure}

In addition, each batch includes a selection vector (SV) \cite{DBLP:conf/damon/NgomMB0LMP21}. The selection vector is a sorted and dense position list containing indices of the rows actually present in the batch (called \emph{active} rows). The selection vector is used to efficiently represent rows excluded from the batch, for example when implementing filtering. Operators access the rows in a batch indirectly via the selection vector, such that iterations over the batch can directly skip over the inactive rows. The BARQ batch layout is illustrated in \cref{fig:batch} with the selection vector in the bottom right. Similar to the actual data (RDF terms), variables are also represented by IDs during execution.

The selection vector is used by several operators, such as \texttt{FILTER}, \texttt{DISTINCT}, and \texttt{MINUS}, to efficiently exclude rows according to different conditions. In the case of vectorized \texttt{FILTER} evaluation, it is sufficient to read the relevant columns, evaluate the filter expression, and, if it evaluates to false, remove the corresponding row index from the selection vector. That helps reusing batches longer during execution instead of copying them in each operator.

We considered using bitmaps as an alternative which tends to work better with SIMD commands. However, research \cite{DBLP:conf/sigmod/BehmPAACDGHJKLL22,DBLP:conf/damon/NgomMB0LMP21} has shown that the selection vector approach works better for complex queries, where batches might contain many inactive rows since operators iterate only over the active rows, not the entire batch.

\subsubsection*{NULLs} Importantly, the set of columns (variables) is fixed for all rows in the batch, i.e., rows are always aligned. This creates a technical challenge related to \texttt{NULL}s. Even though there are no \texttt{NULL}s in the RDF data model, they do appear in query results, for example, when left-outer joins (\texttt{OPTIONAL}s in SPARQL) are used. For a row-based operator, it is not required that all rows in its output have values for the same set of variables. However, when rows with missing values end up in the same batch with full rows, BARQ uses special marker constants to represent \texttt{NULL}s. Those are similar to the special \texttt{NULL} columns in Photon (see Figure 2 in \cite{DBLP:conf/sigmod/BehmPAACDGHJKLL22}). The constants are then treated differently in downstream operators, such as, when checking join conditions or evaluating expressions.

\subsubsection*{Vectorization in Java}
\label{sec:using_the_jvm}

Stardog is primarily a Java-based system (see \cref{sec:architecture}) with many customers running Java 11 that does not have vectorization support at the language or API level\footnote{The Vector API was proposed as a part of JEP 426 in Java 19.}. Nonetheless, the JVM compiler can be highly efficient at compiling code that is amenable to vectorization, i.e., it is able to unroll loops, add SIMD instructions, and more. Prior research \cite{kikolashvili2019design} has shown that it is viable to use Java for vectorized execution code as long as certain features of the JVM are avoided like generics, autoboxing (primitive to object type conversion), or large polymorphic class hierarchies.

Following these observations, we decided that the JVM was going to perform sufficiently well for our use case and implement BARQ operators in Java. This is a pragmatic choice that allows us to reuse  many of our existing components while integrating BARQ into the rest of the system (see \cref{sec:integration}). This particularly applies to the query optimizer, legacy query operators, and the memory management framework, all of which are implemented in Java.

\subsection{The Vectorized Merge Join}
\label{sec:merge_join}

The relational merge join is a fundamental join algorithm in Stardog. In the binary inner equi-join case, it works on two sorted input relations (left and right) and produces an ordered multi-set of rows, matching every pair of rows where values of the join key variable are equal on both sides \cite{DBLP:journals/ibmsj/BlasgenE77}. The merge join improves on the loop join by taking advantage of the input sorted-ness, skipping over non-matching rows on one side if the other side has already produced a row with a greater value of the join variable.

The implementation of BARQ’s vectorized merge join is inspired by the vectorized merge join in CockroachDB \cite{DBLP:conf/sigmod/TaftSMVLGNWBPBR20,CockroachDBVectorizedMJ}. The contribution of BARQ is adding \texttt{skip()} into the vectorized algorithm to take full advantage of Stardog's sorted indexes (see \cref{sec:storage}). To this end, we decompose the classical merge join into three phases: \textit{Probe}, \textit{Build} and \textit{Skip}. 
Given two solution batches (one per input operator) the algorithm runs through the following steps until no more output can be produced:

\begin{enumerate}[label=\arabic*.]
    \item \textbf{Probe:} Determine matching input groups that need to be materialized from the input batches last produced by the left and right operators.
    \begin{enumerate}[label=\alph*)]
        \item A group is defined to be a pair of ranges, where a range is a section of the batch with the same value of the join variable.
        \item The join variable value in both ranges in a group is the same (we sometimes call it the \emph{ordinal} value).
    \end{enumerate}
    
    \item\label{ref:mj_2} \textbf{Build:} Take these groups and materialize them one column at a time:
    \begin{enumerate}[label=\alph*)]
        \item The key to row materialization is that the algorithm only needs to know the group length. It does not need to access any other information.
        \item\label{ref:mj_2b} Each left input value in a column is expanded according to the right range length.
        \item\label{ref:mj_2c} The right column values in a range are repeated according to the left range length.
        \item This essentially computes a column-based cross product, since the algorithm never needs to look at more than one column at a time.
    \end{enumerate}
    
    \item \textbf{Skip:} After materializing results for all groups in the current pair of batches, the algorithm can issue skip calls on the child operators:
    \begin{enumerate}[label=\alph*)]
        \item It issues a \texttt{skip()} call on the operator whose last (greatest) non-matching value of the join variable is less than the last value in the other batch. 
        \item If there is no row with a non-matching value, i.e. all rows were part of an input group, the algorithm simply discards the current batch.
        \item A new input batch is fetched from the operator that the algorithm skipped on, and the process repeats from the probe step (unless the operator terminates, in which case the algorithm finishes, too).
    \end{enumerate}
\end{enumerate}

\begin{figure}[t]
    \centering
    \includegraphics[width=0.48\textwidth]{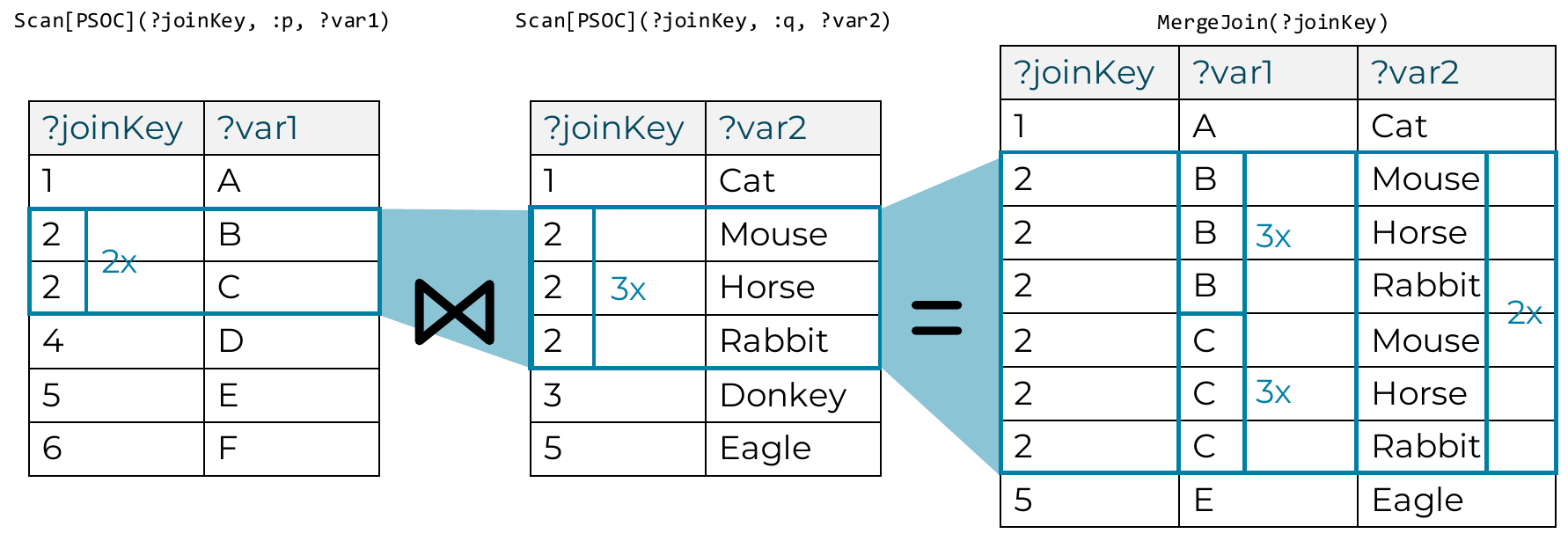}
    \caption{Merge Join: Illustrative example of input ranges and the resulting materialized join output.}
    \label{fig:merge_join}
\end{figure}

\cref{fig:merge_join} exemplifies how the batch-based merge join works. Using full values instead of dictionary-encoded data for clarity, it shows how the group of the join key value \texttt{2} is built in the final cross phase: Each row of the left side (just \texttt{?var1} here) is expanded three times, and the whole range of the right side (just \texttt{?var2} here) is repeated twice.

The key observation is that the algorithm works in a lazy streaming fashion. It materializes the next output batch only when it is requested by the parent operator and it only fetches argument batches that are necessary for producing one output batch. However, it may need to fetch more than one batch from the right operator in case the range for the current join variable value does not fit into a single batch. That can happen because batch size is restricted whereas the max range size is not (as it depends on the data and the query). The algorithm needs to know the exact range size on the right to perform the Build step. To avoid the risk of running out of memory, the algorithm adds these batches into a special collection that can spill off to disk, if necessary (see \cref{sec:mm}). This happens relatively rarely with merge joins in SPARQL (one such notable case are joins on the predicate variable since the number of distinct predicates is typically low and therefore many rows tend to have the same value in the corresponding column).

\subsubsection*{Multiple Join Keys}

At the time of writing, the merge join operator only expects its child operators to provide input sorted by one variable only (the primary join key). However, there are queries where joined patterns share more than one variable, i.e. the join must handle multiple join keys. Since other columns are not guaranteed to be sorted by any value, the join cannot use that property to its advantage. However, the group indexes of the probe phase above are still present, so the join algorithm can leverage them to efficiently perform vectorized equality checks in a single extra pass over the secondary join key columns. This is implemented as an additional \textit{filter} step after the build phase. 

During the first part of that filtering stage, only the join key columns from the right input are materialized in the output batch. This happens as a part of the build phase for the right input in step \ref{ref:mj_2}~\ref{ref:mj_2c}. At that point, the output columns contain the values that must be compared against the left input. For the vectorized comparison, the expansion step is repeated for the left input \ref{ref:mj_2}~\ref{ref:mj_2b} for each joined column. Then the equality condition is checked in one pass over the left input column and the output columns. All rows with mismatches join key values are removed from the output batch using the selection vector (see \cref{sec:selection}).

\subsubsection*{Outer Joins}

Currently, the merge join algorithm in BARQ supports only inner joins. However, it would require only a relatively minor post-processing extension to support left-outer joins (used by Stardog to implement \texttt{OPTIONAL} in SPARQL). 
The left-outer join is defined on a per-row basis, such that if the join condition fails for some row from the left argument, it is added to the output without variable bindings from the right argument. The join condition in \texttt{OPTIONAL}s in SPARQL may also include \texttt{FILTER} expressions\footnote{\url{https://www.w3.org/TR/sparql11-query/\#optionals}}. Thus, in addition to checking equality over the join columns, outer joins should also evaluate SPARQL filters directly on the output batch. That requires tracking the number of output rows for each join group so that if all materialized rows are excluded from the output, a single row with just the left column values remains.

\subsection{Vectorized Streaming Aggregation}
\label{sec:streaming_aggregation}

Similar to SQL, SPARQL 1.1 has rich support for grouping and aggregation\footnote{\url{https://www.w3.org/TR/sparql11-query/\#aggregates}}. Stardog's legacy engine implements general \texttt{GROUP BY} with aggregation functions using a hash-based approach but also offers an optimized implementation when there is a single group variable and the rows to be aggregated come in the order sorted by that variable. This is a pretty common case, see the example query in \cref{lst:aggregation_query}.
The optimizer will pick the merge join to evaluate the graph pattern because there exist indexes storing \texttt{:knows} and \texttt{:interest} triples in the subject order (see \cref{sec:storage}). The results of the join will be sorted by values of the \texttt{?person} variable (the join key). That allows \textit{streaming} aggregation that is much cheaper and has little memory footprint.

Streaming aggregation is particularly suitable for BARQ for two reasons: First, most standard aggregation functions, like  \texttt{count}, \texttt{min}, \texttt{max}, or \texttt{average} are associative. They can be computed over a single column in a batch in a vectorized way and then merged across batches. Second, there is no need to build a hash table using Stardog's memory management layer that is currently row-based (see \cref{sec:mm}).
In addition to grouping, BARQ supports \texttt{DISTINCT} under similar conditions. BARQ's \texttt{DISTINCT} operator uses the \texttt{skip()} method to scroll the input to the next greater value in the sorted order. That is highly efficient for queries with many duplicates. Supporting general hash-based grouping in a vectorized way is left for future work.

\begin{lstlisting}[
%aboveskip=\smallskipamount,
%belowskip=\smallskipamount,
%float=b,
basicstyle=\ttfamily\footnotesize,
language=SPARQL, 
mathescape=true,
morekeywords={GROUP,MAX,COUNT}, commentstyle=\color{gray},
sensitive=true,
firstnumber=0, 
%numbers=left,
numbersep=5pt,
numberstyle=\scriptsize,
frame=none,
caption = {Streaming aggregation: compute the number of unique friends and interest tags for each person}, 
captionpos=b, 
label={lst:aggregation_query},
escapechar=",
columns=flexible,
%xleftmargin=.07\textwidth, 
numberfirstline=false
]
SELECT ?person 
       (COUNT(DISTINCT ?friend) AS ?friends) 
       (COUNT(DISTINCT ?interest) AS ?interests) {
    ?person :knows ?friend .
    ?person :interest ?interest .
}
GROUP BY ?person
\end{lstlisting}

\subsection{Overfetching Problem \& Adaptive Batch Size}
\label{sec:adaptive_batch_size}

One important issue of batch-based processing is that it can lead to reading more data from sorted indexes on disk than is strictly necessary. Consider the following example query that fetches data for a certain product type (\texttt{:ProductType22}). 

\begin{lstlisting}[
%aboveskip=\smallskipamount,
%belowskip=\smallskipamount,
%float=b,
basicstyle=\ttfamily\footnotesize,
language=SPARQL, 
mathescape=true,
morekeywords={BSBM}, commentstyle=\color{gray},
sensitive=true,
firstnumber=0, 
%numbers=left,
numbersep=5pt,
numberstyle=\scriptsize,
frame=none,
%caption = {Query to fetch BSBM product data}, 
%label={lst:bsbm_bgp},
escapechar=",
columns=flexible,
%xleftmargin=.07\textwidth, 
numberfirstline=false
]
SELECT * {
    ?product rdf:type :ProductType22 .
    ?product :productFeature ?feature .
    ?product :producer ?producer .
    ?offer :product ?product .
}
\end{lstlisting}

The query is for the BSBM dataset later used in our evaluation in \cref{sec:experimental_setup} and shows a very common form of a SPARQL query. The query contains a single basic graph pattern (BGP). Stardog evaluates BGPs by joining the solutions of the individual triple patterns and each triple pattern is evaluated using an index scan. Each scan operates on a sorted index and therefore the intermediate results can be merge-joined (see \cref{sec:storage}). This is different from most relational databases where similar data would be stored in a single table and could be read with a single table scan. This leads to the proliferation of joins whose performance is therefore critical.

\begin{figure}[!ht]
\captionsetup{type=lstlisting}
\begin{sublstlisting}{\linewidth}
\begin{lstlisting}[
basicstyle=\ttfamily\tiny,
mathescape=true,
morekeywords={Scan,NaryJoin}, 
commentstyle=\color{gray},
sensitive=true,
numberstyle=\scriptsize,
frame=none,
escapeinside={<@}{@>},
columns=flexible,
numberfirstline=false
]
MergeJoin(?product), results: 2.3M
+- Scan(?product, rdf:type, :ProductType22), results: 5.7K (skip: 3)
+- Scan(?product, :producer, ?producer), results: 11K (skip: 5.5K)
+- Scan(?product, :productFeature, ?feature), <@\textcolor{sd_mediumblue}{\textbf{results: 119K}}@> (skip: 5.5K)
`- Scan(?offer, :product, ?product), results: 120K (skip: 5.5K)
\end{lstlisting}
\caption{Legacy row-based evaluation}
\label{lst:abs_legacy}
\end{sublstlisting}
\begin{sublstlisting}{\linewidth}
\begin{lstlisting}[
basicstyle=\ttfamily\tiny,
mathescape=true,
morekeywords={Scan,NaryJoin}, 
commentstyle=\color{gray},
sensitive=true,
numberstyle=\scriptsize,
frame=none,
escapeinside={<@}{@>},
columns=flexible,
numberfirstline=false
]
MergeJoin(?product), results: 2.3M (next: 4.5K), wall time: (4.5%)
+- Scan(?product, rdf:type, :ProductType22), results: 5.7K (next: 13), batched
+- Scan(?product, :producer, ?producer), results: 260K (next: 509, skip: 502), batched
+- Scan(?product, :productFeature, ?feature), <@\textcolor{sd_mediumblue}{\textbf{results: 2.0M}}@>  (next: 3.8K, skip: 3.7K), batched
`- Scan(?offer, :product, ?product), results: 1.9M (next: 3.7K, skip: 3.6K), batched
\end{lstlisting}
\caption{BARQ, fixed batch size for each scan operator}
\label{lst:abs_barq_no_abs}
\end{sublstlisting}
\begin{sublstlisting}{\linewidth}
\begin{lstlisting}[
basicstyle=\ttfamily\tiny,
mathescape=true,
morekeywords={Scan,NaryJoin}, 
commentstyle=\color{gray},
sensitive=true,
numberstyle=\scriptsize,
frame=none,
escapeinside={<@}{@>},
columns=flexible,
numberfirstline=false
]
MergeJoin(?product), results: 2.3M (next: 4.5K), batched
+- Scan(?product, rdf:type, :ProductType22), results: 5.7K (next: 58), batched
+- Scan(?product, :producer, ?producer), results: 40K (next: 5.1K, skip: 5.0K), batched
+- Scan(?product, :productFeature, ?feature), <@\textcolor{sd_mediumblue}{\textbf{results: 139K}}@> (next: 17K, skip: 5.5K), batched
`- Scan(?offer, :product, ?product), results: 146K (next: 17K, skip: 5.5K), batched
\end{lstlisting}
\caption{BARQ, adaptive batch size}
\label{lst:abs_barq}
\end{sublstlisting}
\caption{Adaptive Batch Size Example}
\label{lst:adaptive_batch_size}
\end{figure}

\Cref{lst:abs_legacy} shows a simplified Stardog profiler\footref{fn:profiler} output for the join that evaluates the BGP in the example query. The logical plan uses an n-ary merge join which, in the case of BARQ, is currently translated into a left-deep binary join tree. The most selective triple pattern is \texttt{?product rdf:type :ProductType22}: with the corresponding \texttt{Scan} operator producing only \texttt{5.7K} rows. The merge join algorithm repeatedly uses the \texttt{skip()} method on each argument operator to jump to the next row that has a matching value for the join key variable (\texttt{?product}) (or the next greater value). By doing so, it greatly reduces the number of unnecessary disk reads, such as for producers or product features for products of types other than \texttt{:ProductType22}. Non-matching triples are skipped over directly at the storage layer using RocksDB \texttt{seek} API.

This proved to be a challenge for BARQ since fetching a \textit{fixed} number of rows per \texttt{next()} call could lead to a lot of unnecessary disk IO. When a \texttt{Scan} operator in BARQ reads batches from the storage, it has no information on how the parent operator will use the batch. In the case of merge joins, which heavily use skipping, it is quite likely that most of the batch will be discarded. For example, consider the \texttt{:productFeature} scan in \Cref{lst:abs_barq_no_abs}. It reads \texttt{2.0M} triples from disk, which is an order of magnitude more than necessary, compared with the row-based scan in \Cref{lst:abs_legacy}, which reads \texttt{119K} triples. This is because it batches features for products of many types, most of which do not satisfy the join condition.

We call this problem \textit{overfetching}. It leads to increased disk IO, memory consumption, and processing time, thus partly negating the benefits of vectorized joins. It mostly affects OLTP-style queries, which used to be the sweet spot for Stardog's legacy query engine.

To this end, BARQ implements a novel \textit{adaptive batch sizing} technique. Based on the pattern of \texttt{next()/skip()/reset()} calls that an operator receives from its parent, it adapts how many rows are going to be produced in the next batch. Many common operators in Stardog exhibit characteristic read patterns. Whereas merge joins heavily use \texttt{skip()}, the \texttt{Sort} operator materializes the entire output of its argument operator with only \texttt{next()} calls, thus the argument will quickly increase the batch size for each \texttt{next()} call up to the defined limit. This is also true for other pipeline-breakers, like the hash join, non-streaming \texttt{GROUP BY}, and \texttt{ORDER BY}.

\Cref{lst:abs_barq} shows the effectiveness of batch size adaptation. The output for each scan is greatly reduced and is much closer to the row-based merge join than to the naive BARQ evaluation with a fixed batch size. Some unnecessary reads appear unavoidable and we discuss their impact further in \cref{sec:experimental_results}.

The batch size tends to be small for the leaf operators, most of which tend to be index scans producing sorted output, and then increases as the batches travel up the operator tree. This is the result of BARQ adapting the number of produced rows per operator based on different factors. One is the properties of the underlying data. Joins that process batches with multiple rows per join key value (like in \cref{lst:adaptive_batch_size}) tend to produce more results than their inputs. Another factor is pipeline-breaking operators. They become more common towards the top of the operator tree since intermediate results need to be re-sorted or added into a hash table.

\section{Integrating BARQ into Stardog}
\label{sec:integration}
Developing a new batch-based query executor (or a prototype thereof) is already hard but the base principles are well-known at this point. At the same time, integrating a new batch-based executor into an existing database system, which has been built on different principles, is a different challenge. In the ideal scenario, a new executor would be a drop-in replacement for the legacy, row-based one. However, in our experience described below, that is rather unlikely.

\subsection{BARQ: a Drop-in Replacement or Not?}
\label{sec:barq_replacement}

Stardog has well-defined API boundaries between the components, particularly, between the optimizer, translator, and executor. The translator receives an optimized logical plan and translates it into a tree of \texttt{Operator} instances, executable by calling \texttt{next()} from the top. Neither the translator nor the server’s layer that subsumes \texttt{Operator}’s data have any visibility into their internals. That should — in theory — make it possible to replace one executor with another by plugging a different translation layer\footnote{Even though it requires batch-to-row adapters to integrate with the tuple-at-a-time Operator API, it does not, by itself, preclude drop-in replacement provided that every operator in the plan is batch-based (so the adapter is only used at the top). That requirement is, however, challenging, as explained below.}. Yet, there are obstacles that make it very difficult for mature systems:

The first problem is the sheer number of different operators required to implement an expressive query language. Apart from the BGP evaluation that (at the minimum) requires traversal or join algorithms, SPARQL has plenty of common relational operators, such as filters (selections), projections, \texttt{BIND} (extensions), unions, reachability operators (property paths), aggregations, etc. There is at least one \texttt{Operator} implementation for each. However, similar to many mature systems, Stardog has multiple algorithms for performance-critical operators, such as joins, anti-joins (\texttt{MINUS}), or aggregation, allowing the optimizer to pick the best one for each point in the query plan. Re-implementing all of them to process data in batches would not be a good way to bring BARQ to production in a reasonable time frame.

Second, not all relational operators are can be vectorized to the same degree or with the same effort \cite{DBLP:conf/damon/ZukowskiNB08}. While batch-based evaluation of joins or filters has been thoroughly studied, this is less true, for example, for recursive operators, like property paths that are used for graph reachability queries. In addition, Stardog supports custom SPARQL extensions through the \texttt{SERVICE} mechanism (originally included in SPARQL for query federation purposes\footnote{\url{https://www.w3.org/TR/sparql11-federated-query/}}). It uses services for functionalities like full-text keyword search, spatial search, regression, and classification models, as well as supporting pluggable user-defined services. It is unreasonable to require that all peripheral operators become batch-based in the same release as the core operators.

Third, it is not clear if batch-based query execution will ever perform better on \texttt{every} query in the foreseeable future. Particularly, there will always be selective, IO-bound queries where batch reads from disk tend to lead to overfetching (see \cref{sec:adaptive_batch_size}). Furthermore, there could be hybrid queries with both CPU-bound and IO-bound sub-queries so there is an appeal for combining both execution strategies even within a single query.

Finally, Stardog's memory management layer integrates tightly with the query engine (see \cref{sec:mm}). The layer serializes intermediate query results into byte arrays to avoid Java object overhead and support spilling to disk. The data format (both in memory and on disk) had been optimized for the legacy per-row operators and requires adapters to work for the batch-based operators. Re-designing the memory management layer to natively support batch-based and column-oriented processing would be another large-scale project comparable in scope with the query executor itself.

Given these considerations, we have decided to introduce BARQ into Stardog's query execution engine in a gradual manner, incrementally replacing the legacy per-tuple operators with the new batch-based operators. The approach mitigates the risks of bringing a new core component into production, decreases the delivery time, and ensures that at every stage the query engine satisfies all quality assurance constraints (e.g., passes all correctness and performance regression tests). At the same time, it is not without its own challenges that we discuss next.

\subsection{Integration Challenges}
\label{sec:integration_challenges}

The co-existence of two different query executors is a double-edged sword (although not unheard of, see \cref{sec:related_work}). It allowed us to bring BARQ into production faster while keeping the system stable. It also enables the query engine to choose the right execution approach for each query (or even a subquery). At the same time, it raises the following challenges:
\begin{itemize}
    \item \textbf{\hyperref[sec:interoperability]{Interoperability}:} If different parts of the query plan are executed with different executors, they should be combined with the minimal overhead.
    \item \textbf{\hyperref[sec:selection]{Selection}:} How to decide which executor should be used for a query (or a particular part of the query)?
    \item \textbf{\hyperref[sec:component_isolation]{Component Isolation}:} To what extent should the rest of the system be oblivious to how a physical plan is executed? In particular, should the optimizer (and the cost model, as a part of it) be aware of the executor differences or not?
\end{itemize}

\subsubsection*{Interoperability}
\label{sec:interoperability}

\begin{figure}
    \centering
    \includegraphics[width=0.37\textwidth]{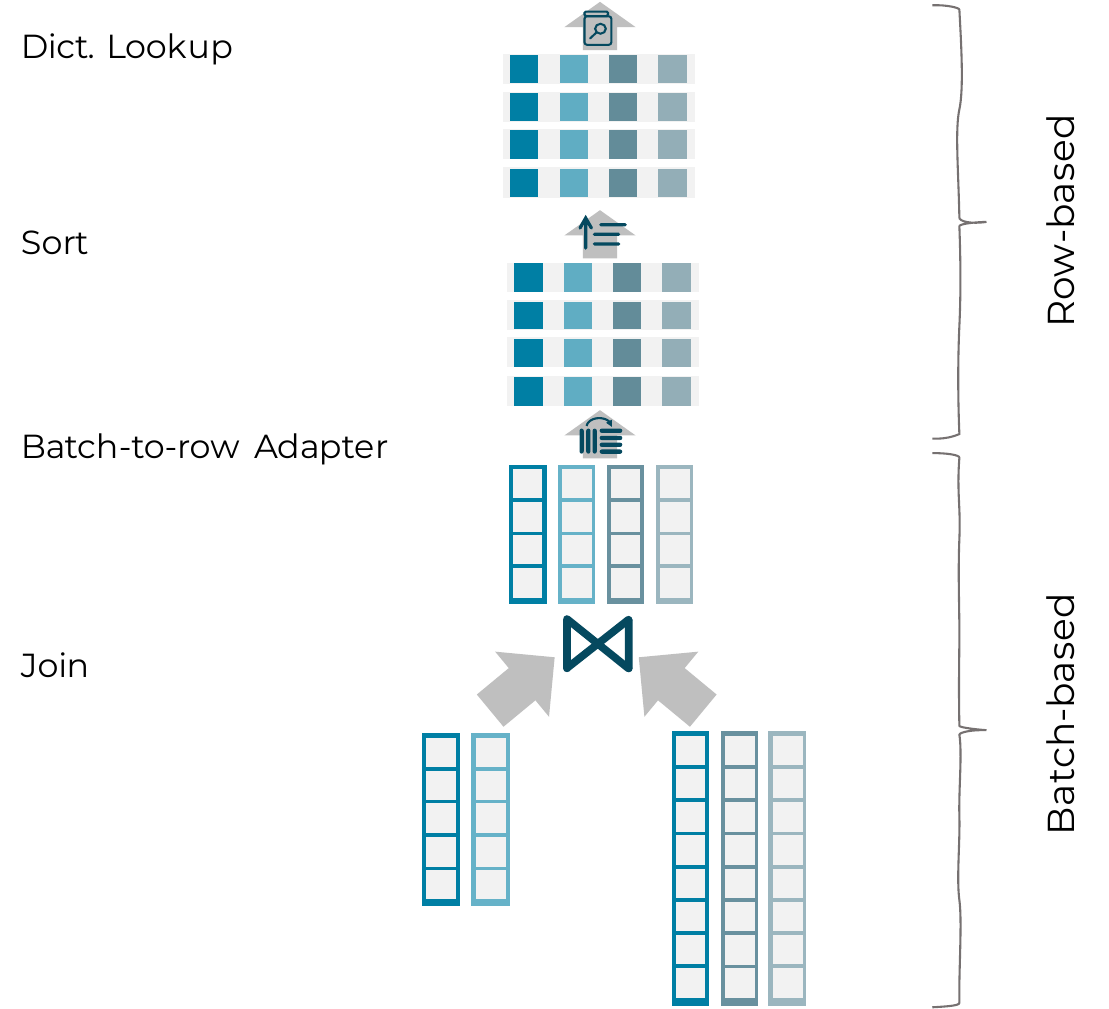}
    \caption{Example of batch-to-row operator adapters}
    \label{fig:execution_example}
\end{figure}

We take advantage of the fact that it is straightforward to insert operators for any kind of auxiliary tasks in the Volcano model (one example is the exchange operators for parallelism \cite{10.1145/93605.98720}). We use batch-to-row adapters to pivot batches to rows so that per-row operators can consume results produced by BARQ operators.
The adapter implementation is rather straightforward. Their overhead turns out to be negligible in our experiments because they are copy-free: a batch of rows can be immediately processed as an array of rows. Also, the number of integration points in any given executable plan is usually low. 
Most operators near the bottom of the plan tend to be batch-based because BARQ supports scans and merge joins (most joins over scans are merge joins due to the sorted indexes, see \cref{sec:storage}). If an unsupported operator occurs in some place in the plan, batch-to-row operators are injected below and its input data is transformed to rows (see \cref{fig:execution_example} for an example where the conversion happens between a BARQ join and a row-based sort operator). All parent operators can then use per-row algorithms.

It is also possible to convert intermediate results in the other direction, from rows to batches. That can happen if a legacy operator produces many results that are going to be consumed by an operator for which a BARQ algorithm exists. One example could be a join over a property path. There is an explicit row-to-batch adapter but typically such conversion happens at a pipeline-breaking point, e.g. where the data is sorted or inserted into a hash table. Whenever data could be spilled to disk, both legacy and BARQ operators serialize the data into memory blocks in the same binary format. That means that the output of a per-row operator, once sorted, can be read back as a stream of batches and sent directly to batch-based operators.

\subsubsection*{Selection: BARQ or Row-Based Operators}
\label{sec:selection}

It is currently the query plan translator’s job to decide if a particular operator should use a BARQ implementation or a row-based implementation. The decision is based on several factors: 

\begin{itemize}
    \item whether there is a BARQ-supported implementation for the operator,
    \item whether the child operators generate batches or rows, and
    \item how much data the operator is expected to process internally compared with its child operators.
\end{itemize}

If there is a BARQ-native implementation for an operator \textit{and} all child operators use BARQ, then the operator uses BARQ. If there is no BARQ implementation, then the decision is similarly straightforward. The important decision point is when the operator can use both execution algorithms and some of its arguments cannot use BARQ (so their \texttt{next()} methods return single rows). In that case, any potential performance gains from using batch-based processing have to be weighed against the risk of overfetching (i.e., reading more data than necessary due to batching, see \cref{sec:adaptive_batch_size}). Currently, Stardog’s translator does so for merge joins that are expected to generate more results than any of their arguments. For such joins most of the work typically happens in main memory. The decision is cost-based and depends on cardinality estimations.

\subsubsection*{Component Isolation: Two Executors and One Optimizer}
\label{sec:component_isolation}

As soon as it became clear that Stardog is going to live with two execution models for a while, the question arose: shall the optimizer be aware of the differences between the two? The relevant parts of the optimizer here are the cost model and the join order selection. Other optimization rules and the cardinality estimation framework are largely oblivious.

In an attempt to contain the BARQ project within the query executor component of the system, we first attempted to keep the optimizer unaware. The optimizer would produce an execution plan based on a single cost model regardless of whether BARQ is even enabled, and the translator would then make the selection. That was based on the assumption that the cost of each plan is independent of the execution strategy. Of course, that assumption only holds up to certain limits but, first, so are many other simplifying assumptions that the cost model must make and, second, it is not the absolute value of the cost that matters but the relative order of costs for alternative execution plans (particularly, join trees).

However imperfect that assumption is, it allows us to avoid maintaining two execution-specific cost models. That would have been very expensive considering that every change to the cost model requires comprehensive performance testing on a range of internal benchmarks. It is not uncommon that even a minor cost model change makes some queries faster but then a few slower, and in general one can write a separate paper on how to decide if the change is a net positive. It would have been unwise to add these complications to the already ambitious BARQ project.

\begin{lstfloat}[t]
\begin{lstlisting}[
%aboveskip=\smallskipamount,
%belowskip=\smallskipamount,
basicstyle=\ttfamily\footnotesize,
%language=SPARQL, 
mathescape=true,
morekeywords={Scan,MergeJoin,Scan,Sort,Group,Filter,BindJoin,HashJoin,Block},
commentstyle=\color{gray},
sensitive=true,
firstnumber=0, 
numbersep=5pt,
label={lst:different_plan},
caption={Legacy Engine Query Plan for LSQB Q6},
numberstyle=\scriptsize,
frame=none,
escapeinside={<@}{@>},
numberfirstline=false
]
Group(aggregates=[(COUNT(*) AS ?count)])
`- BindJoin(?person3)
   +- Scan(?person3, :interest, ?tag)
   `- Filter(?person1 != ?person3)
      `- HashJoin(?person2)
         +- Scan[POSC](?person1, :knows, ?person2)
            `- MergeJoin(?person3)
               +- Scan[POSC](?person2, :knows, ?person3)
               `- Block(sortedBy=?person3)
\end{lstlisting}
\end{lstfloat}

Still, sometimes performance advantage of BARQ is just too large to ignore. One such case is merge joins that generate substantially more results than either of the child operators. That means that most of the query processing happens inside the join itself, not the child operators, and it is CPU-bound. The cost model makes limited provisions for that specific case, assigning a lower cost if the arguments are BARQ-supported. That sometimes yields different execution plans for BARQ. \Cref{lst:different_plan} shows the query plan of the legacy engine for the query of our motivating example (the BARQ plan is shown above in \cref{lst:motivation}).

In this case, the legacy merge join has high interpretation overhead and the optimizer decides to use a bind join\footnote{Similar to a block-nested loop join, the block-based bind join in Stardog retrieves blocks of tuples ($\approx$ 1K tuples) from its left-hand side and pushes the block of tuples into the right-hand side. The right-hand side is evaluated for each pushed block to produce the join results.} on top of the plan. It effectively leads to computing the two-hop closure of the \texttt{:knows} graph for a block of interest tags at a time. That is more efficient for the legacy engine than the plan in shown \Cref{lst:motivation}. However, with the BARQ execution engine, the query plan that only uses merge joins and sort operators is the most efficient, so the cost model was tuned to make it cheaper.

\section{Evaluation}
\label{sec:evaluation}
We present experiments on different workloads to compare the query execution performance of BARQ to the legacy execution engine. The goal of the experiments is to answer two questions: 
\begin{enumerate}[label=\textsf{QN\arabic*}]
    \item \label{ref:q1} To which degree is BARQ more efficient in processing CPU-bound queries?
    \item \label{ref:q2} How does BARQ perform on IO-bound queries with relatively little data in memory?
\end{enumerate}

We use the LSQB benchmark \cite{DBLP:conf/sigmod/MhedhbiLKWS21} for the CPU-bound and the BSBM benchmark \cite{DBLP:journals/ijswis/BizerS09} for OLTP and IO-bound performance assessments. LSQB has been specifically designed to test (in-memory) join performance on graph queries that are not constrained to specific nodes. For BSBM, which is a benchmark built around an e-commerce scenario, we use two different use cases: The Explore use case where all queries use selective criteria and look up only a few rows on disk and the Business Intelligence (BI) use case where queries read substantially more data to answer business intelligence queries but disk access still dominates in-memory processing.

\begin{figure}
  \begin{subfigure}[t]{1.\linewidth}
    \centering\includegraphics[width=0.95\textwidth]{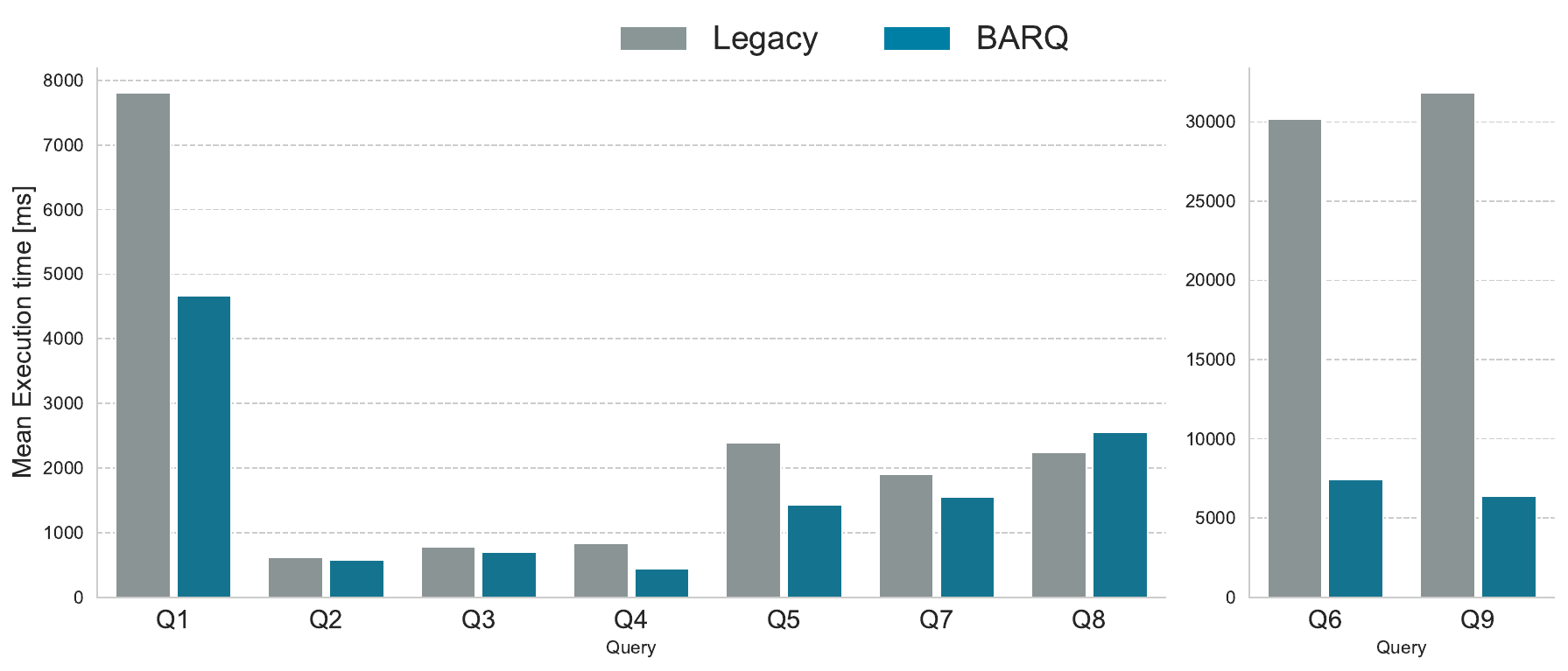}
    \caption{LSQB}
    \label{fig:lsqb_results}
  \end{subfigure}
  \hfill
  \begin{subfigure}[t]{1.\linewidth}
    \centering\includegraphics[width=0.95\textwidth]{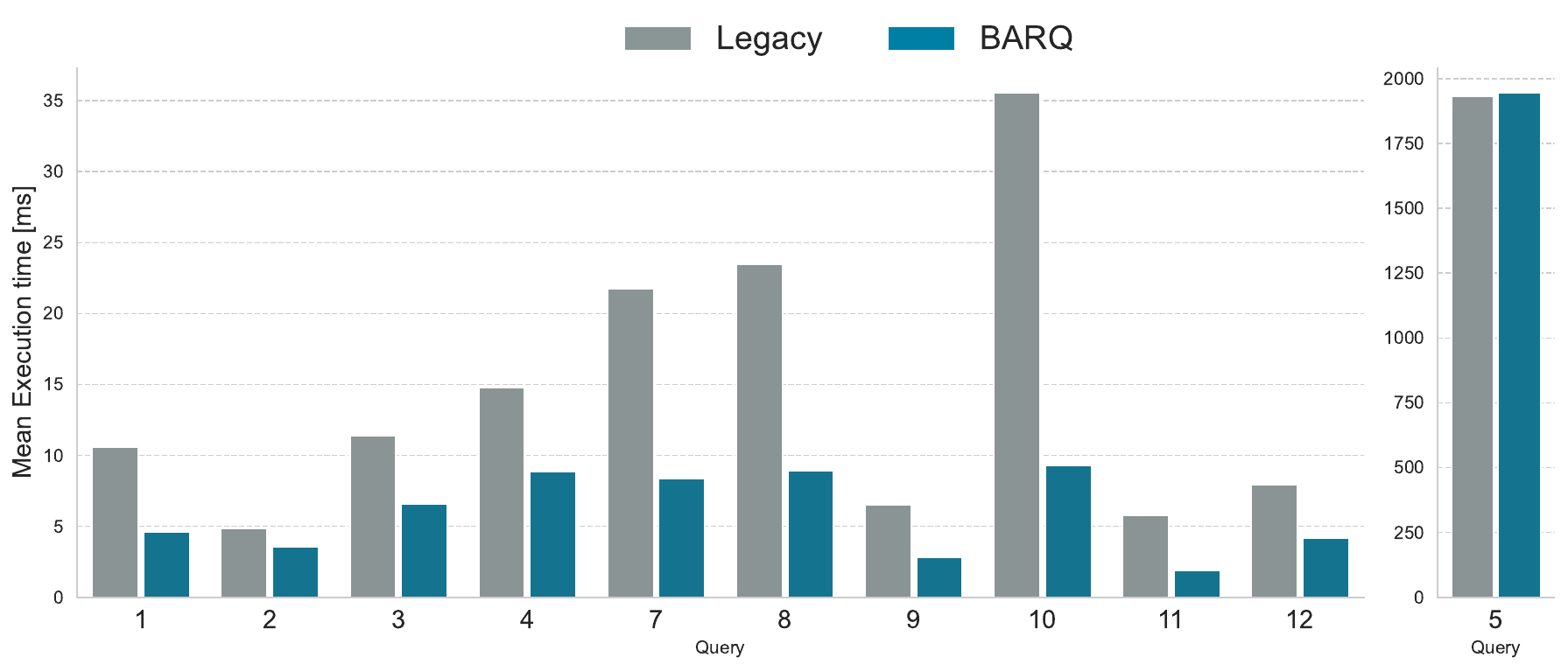}
    \caption{BSBM Explore}
    \label{fig:bsbm_explore_results}
  \end{subfigure}  
  \hfill
  \begin{subfigure}[t]{1.\linewidth}
    \centering\includegraphics[width=0.95\textwidth]{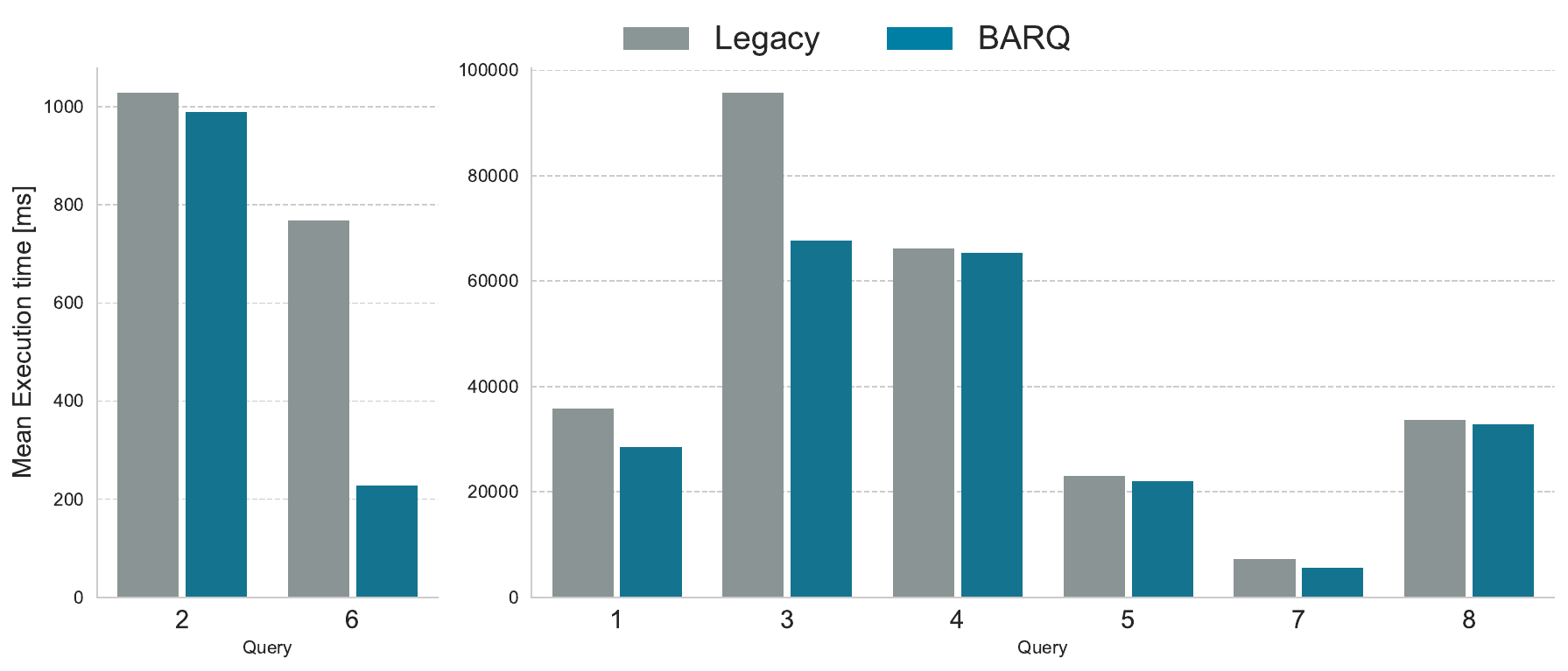}
    \caption{BSBM Business Intelligence}
    \label{fig:bsbm_bi_results}
  \end{subfigure}    
\caption{Benchmarking BARQ against the legacy engine}  
\label{fig:experimental_results}
\end{figure}

\subsection{Experimental Setup}
\label{sec:experimental_setup}

We use a memory-optimized AWS EC2 \texttt{r4.2xlarge}\footnote{\url{https://aws.amazon.com/ec2/instance-types/}}  with 8 vCPUs on a 2.3 GHz Intel Xeon Scalable Processor and 61 GB of RAM and ran Stardog server version 11\footnote{to be released in January 2025}. We allow the JVM to use up to 16 GB of heap memory and 30 GB off-heap memory. The system under test is Stardog 11.1 pre-release (scheduled for release on 05/2025).

For the LSQB benchmark, we use a scale factor of $0.3$ which leads to a dataset with \num{7363374} RDF triples. Note that, regardless of the small dataset size, the LSQB queries (specifically Q6 and Q9) stress the system due to the extremely large number of intermediate results (>\texttt{250M}) that need to be processed. That creates a clear computational bottleneck that allows us to compare the performance of row-based and vectorized query operators on CPU-bound queries. For each query, we execute 3 warm-up runs and average the execution time over 6 test runs. 

For BSBM, we use the data generator\footnote{\url{http://wbsg.informatik.uni-mannheim.de/bizer/berlinsparqlbenchmark/spec/index.html}} with a scale factor of \num{2848000} (number of product instances in the data) to obtain a dataset with \num{999700717} RDF triples. We use the official test driver to run the Explore\footnote{\url{http://wbsg.informatik.uni-mannheim.de/bizer/berlinsparqlbenchmark/spec/ExploreUseCase/index.html}} use case with 100 warm-up runs and 25 test runs, as well as the Business Intelligence\footnote{\url{http://wbsg.informatik.uni-mannheim.de/bizer/berlinsparqlbenchmark/spec/BusinessIntelligenceUseCase/index.html}} (BI) use case with 10 warm-up runs and 25 test runs. Note that the benchmark driver uses query templates which are instantiated by replacing placeholders with random constants (e.g., products or product types). For simplicity in the result discussion, we use the term “query” instead and present the aggregated results across all query template instances. For both BSBM use cases, the query execution time (aQET) is averaged over test runs and presented as reported by the BSBM driver. 

For all benchmarks, the measured query execution time includes query optimization time but since Stardog uses the same optimization pipeline and largely the same cost model (c.f., \cref{sec:component_isolation}), any performance differences are attributable to the execution engines.

\subsection{Experimental Results}
\label{sec:experimental_results}

The results for all benchmarks are shown in \cref{fig:experimental_results}. For the sake of readability, in all figures we split the queries into two categories: easy and hard, using different y-axis scales. To address \ref{ref:q1}, we first discuss the results for the CPU-heavy LSQB benchmark.

\subsubsection*{LSQB}

The results shown in \cref{fig:lsqb_results} demonstrate an improvement in the mean query execution time for all queries. The total query throughput for the whole benchmark is 3.4 times higher with BARQ than with the legacy engine. The greatest absolute improvement is observed for the long-running queries Q6 and Q9 with an average time reduction of \qty{27}{\second} (\qty{83}{\percent}) and \qty{33}{\second} (\qty{82}{\percent}), respectively. Both queries are similar: Q9 just adds a \texttt{FILTER NOT EXISTS} condition to eliminate triangular graph patterns (Stardog uses the \texttt{MINUS} operator, akin to SQL’s anti-join, to evaluate it for this query). 

The BARQ profiler output for the full version of Q6 with bi-directional \texttt{:knows}-scans is shown in \Cref{lst:lsq_q6_profile} and demonstrates how fast vectorized merge joins are: The top merge join generates $288.1M$ rows using $563K$ batches while accounting for just over \qty{10}{\percent} of the total execution time (\qty{37}{\percent} if \texttt{Sort} time is included) \footnote{The rest is due to the operators that have not yet been fully vectorized, e.g. filters}. For comparison, the legacy engine spends more than \qty{70}{\percent} of the total time executing joins. It is worth noting that $563$K batches means that the average batch size in this case is $288M / 563K = 506$, which is very close to the max allowed size of $512$. In other words, the adaptive batch sizer realizes that the query is CPU-bound, not IO-bound, so favors larger batches. Finally, generating $563K$ batches instead of $285M$ individual tuples means that the top level \texttt{COUNT(*)} operator is also faster: its share is reduced from \qty{20}{\percent} to \qty{9.2}{\percent} because of the lower interpretation overhead (fewer \texttt{next()} calls to make).

\begin{lstlisting}[
%aboveskip=\smallskipamount,
%belowskip=\smallskipamount,
basicstyle=\ttfamily\footnotesize,
%language=SPARQL, 
float=*, 
%multicols=2, 
mathescape=true,
morekeywords={Scan,MergeJoin,Scan,Sort,Group,Filter,BindJoin,HashJoin,Block,Union},
commentstyle=\color{gray},
sensitive=true,
firstnumber=0, 
numbersep=5pt,
label={lst:lsq_q6_profile},
caption={LSQB Q6: BARQ Query Profile},
numberstyle=\scriptsize,
frame=none,
escapeinside={<@}{@>},
numberfirstline=false
]
Group(aggregates=[(COUNT(*) AS ?count)]), results: 1 (next: 2), wall time: 9.2%, batched
`- Filter(?person1 != ?person3), results: 285.5M (next: 563K), wall time: 49.4%, batched
   `- MergeJoin(?person2), results: 288.1M (next: 563K), wall time: 10.8%, batched
      +- Union, results: 114K (next: 245), wall time: (0.1%), batched
      |  +- Scan(?person1, :knows, ?person2), results: 57K (next: 460), wall time: 0.3%, batched
      |  `- Scan(?person2, :knows, ?person1), results: 57K (next: 460), wall time: 0.3%, batched
      `- Sort(?person2), memory: {total=100M (96.2%)}, results: 2.6M (next: 5.1K, skip: 1), wall time: 27.2%, batched
         `- MergeJoin(?person3), results: 2.6M (next: 5.2K), wall time: 1.3%, batched
            +- Union, results: 114K (next: 245), wall time: 0.1%, batched
            |  +- Scan(?person2, :knows, ?person3), results: 57K (next: 460), wall time: 0.3%, batched
            |  `- Scan(?person3, :knows, ?person2), results: 57K (next: 460), wall time: 0.3%, batched
            `- Scan(?person3, :interest, ?tag), results: 88K (next: 1.4K, skip: 66), wall time: 0.6%, batched
\end{lstlisting}

The performance of some faster-running queries is also substantially improved with BARQ for similar reasons. One example is Q3 that also requires triangular joins on the \texttt{:knows} predicate but includes other parts that restrict the number of intermediate results. BARQ executes it 6 times faster on average.
Next, we take a closer look at the BSBM benchmarks where disk IO plays a larger role.

\subsubsection*{BSBM Explore}

We first discuss the results of the Explore use case shown in \cref{fig:bsbm_explore_results}. They show that BARQ slightly outperforms the legacy engine in all queries except query 5. The overall improvement is negligible with a mean/median reduction of \qty{3}{\milli\second}/\qty{5}{\milli\second} in the execution time across all queries. However, this is the query workload that the row-based engine has been specifically designed for and BARQ has not, but its performance is nonetheless competitive. The results demonstrate the effectiveness of the adaptive batch sizing technique that mitigates the overfetching problem. With the technique turned off, BARQ's query throughput on BSBM Explore is about \qty{33}{\percent} lower.

\subsubsection*{BSBM Business Intelligence}

Finally, we discuss the results for the BI use case that comprises 8 analytical queries as shown in \cref{fig:bsbm_bi_results}. BARQ outperforms the legacy engine on the majority of queries and it has \qty{9.1}{\percent}  higher throughput on the query mix. The largest improvement is observed for query 3 whose execution time is reduced by almost \qty{41}{\percent}. This is a query where the merge joins account for the largest percentage of the execution time and therefore it benefits most from batch-based execution. The second longest-running query is query 4 and currently BARQ shows about the same performance. The main reasons are the lack of support for vectorized aggregation over unsorted inputs and overfetching (even with adaptive batches). With a fixed batch size, BARQ loses about \qty{44}{\percent} of its overall query throughput. It is more than for BSBM Explore and suggests that overfetching is a bigger issue for disk-bound queries.

\subsubsection*{Summary}
\label{sec:summary}

Summarizing the experimental evaluation of BARQ, the results of the LSQB benchmark show that BARQ significantly improves query execution time for CPU-bound queries (\ref{ref:q1}). Moreover, the results show that BARQ performance is on par with the legacy execution engine for IO-bound queries (\ref{ref:q2}). The former demonstrates the benefits of the Vector Volcano approach implemented by BARQ where tuples in each batch can be processed in tight loops with lower interpretation overhead. The latter demonstrates the effectiveness of the adaptive batch sizing technique. Even for the BSBM Explore use case that consists of OLTP-style point lookup queries with selective constants, which process only a few intermediate results, BARQ is able to outperform the legacy engine. This was the key factor enabling us to make BARQ the default query executor in Stardog version 10.2 in September 2024.

Overall, BARQ improvements on CPU-bound queries are unsurprising; the experiments confirm that batch-based processing is far superior when dealing with many intermediate results. At the same time, understanding why BARQ is somewhat faster or somewhat slower than the tuple-at-a-time engine on OLTP queries is more complex. The eventual performance ratio is due to a combination of several interacting factors. For example, BARQ enables higher join and filtering throughput but often reads more data from disk (\hyperref[sec:adaptive_batch_size]{overfetching}). The latter, however, is partly mitigated by the adaptive batch sizing technique. In general, deciding whether one should further optimize a batch-based processing engine for selective queries or use it in combination with tuple-at-a-time operators is not always straightforward.

\section{Related Work}
\label{sec:related_work}
Several commercial and academic vectorized query engines have been developed over the last two decades. While the majority of these engines were developed in the realm of relational databases, many of the approaches are directly relevant to Stardog’s SPARQL engine. The vectorized query engine model was first introduced in MonetDB/X100 \cite{DBLP:conf/cidr/BonczZN05}. Similar to BARQ, the X100 execution engine is a Volcano-style engine which processes chunks of data (vectors) in columnar format. Photon \cite{DBLP:conf/sigmod/BehmPAACDGHJKLL22}, a commercial query engine developed by Databricks, is also a vectorized engine integrated into the Apache Spark runtime. Photon operates on batches of values, called \textit{column vectors}, which also use a columnar layout. Inspired by Photon, BARQ employs \textit{selection vectors}, to store the indexes of active rows in a batch. The vectorized merge join implementation in BARQ is particularly inspired by the merge join algorithm in CockroachDB \cite{DBLP:conf/sigmod/TaftSMVLGNWBPBR20}, which is a commercial SQL database. BARQ follows the same base principles and complements the \texttt{skip()} API to better leverage the sorted indexes supported by Stardog's storage engine.

In the space of graph databases, particularly relevant is Blazegraph\footnote{\url{https://blazegraph.com}}, which is an RDF triple store\footnote{The company behind Blazegraph has been acquired by Amazon that later released its cloud graph database offering called Amazon Neptune.} that also implements a vectorized query engine and processes solutions in batches. In contrast to BARQ, however, the engine does not use a columnar format to exchange data between operators. Newer graph databases products, such as the embedded system called Kùzu \cite{DBLP:conf/cidr/JinFCLS23}, use a vectorized query processor, too, but in addition support the so-called \texttt{factorization} technique to compactly represent join outputs (this is one possible future direction for BARQ). Such systems, by virtue of being new, typically do not need to integrate multiple engines or deal with legacy components. The popular property graph database Neo4j, however, combines three different execution engines (called \emph{Cypher query runtimes}) partly because different engines perform best on different queries\footnote{\url{https://neo4j.com/docs/cypher-manual/current/planning-and-tuning/runtimes/concepts/}}. This observation matches our experience with BARQ reported in this work.

Finally, there are open-source modular query engine systems like Apache Data Fusion \cite{DBLP:conf/sigmod/LambSHCKHS24} and Velox \cite{DBLP:journals/pvldb/PedreiraEBWSPHC22} developed by Meta. They do use technologies that are directly relevant to BARQ, such as Apache Arrow\footnote{\url{https://arrow.apache.org/}} (a general-purpose columnar data format for in-memory computations). These systems can be used as building blocks to assemble new databases. At the same time, they have been designed for relational, strongly typed data with a rigid schema, and the complexity of extending them to implement a graph query language over (possibly schemaless) RDF data is comparable to building BARQ from scratch.

\section{Conclusion}
\label{sec:conclusion}
Integrating a different computational approach into a mature query engine is challenging and, assuming limited engineering resources, often requires making compromises. The key to success is a clear path consisting of many incremental steps, each of which takes the system from one usable state to the next, improved state. Otherwise, the project runs into the risk of substantial delays and losing its focus on the original problem.
This paper describes the development of BARQ, its integration into Stardog, and the lessons we have learned along the way. It was fairly clear from the beginning that batch-based query processing would enable higher throughput on analytical query workloads. Our evaluation confirms that. The main lessons, however, are around keeping the BARQ project focused on the execution part of the query processing pipeline, integrating it with other components in a pragmatic way, and mitigating the specific issues with handling disk-bound and selective queries. The latter requires novel solutions, such as adaptive batch sizing.

The BARQ project was started in August 2023, beta-released in Stardog 10.1 in May 2024, and finally released as the default engine in Stardog 10.2 in September 2024. At the same time, the work is far from over. The downside of restricting the project to vectorizing query operators is that other components, which have been developed for the legacy engine, do sometimes get in the way. Particularly, this applies to the cost model, the memory management framework, and the mapping dictionary. It is our intention to gradually extend the batch-based paradigm to those components as well, for example, by migrating Stardog’s managed collections, such as hash tables or sorted arrays, to a modern columnar data format and by adding a vectorized API for both direct and inverse dictionary look-ups. Once the dictionary supports vectorized look-ups, BARQ can be further extended to cover the virtualization layer since it heavily relies on the dictionary to encode all intermediate query results coming from remote endpoints.

\bibliographystyle{ACM-Reference-Format}
\bibliography{bib}


\begin{thebibliography}{24}


\ifx \showCODEN    \undefined \def \showCODEN     #1{\unskip}     \fi
\ifx \showDOI      \undefined \def \showDOI       #1{#1}\fi
\ifx \showISBNx    \undefined \def \showISBNx     #1{\unskip}     \fi
\ifx \showISBNxiii \undefined \def \showISBNxiii  #1{\unskip}     \fi
\ifx \showISSN     \undefined \def \showISSN      #1{\unskip}     \fi
\ifx \showLCCN     \undefined \def \showLCCN      #1{\unskip}     \fi
\ifx \shownote     \undefined \def \shownote      #1{#1}          \fi
\ifx \showarticletitle \undefined \def \showarticletitle #1{#1}   \fi
\ifx \showURL      \undefined \def \showURL       {\relax}        \fi
\providecommand\bibfield[2]{#2}
\providecommand\bibinfo[2]{#2}
\providecommand\natexlab[1]{#1}
\providecommand\showeprint[2][]{arXiv:#2}

\bibitem[Armbrust et~al\mbox{.}(2015)]%
        {DBLP:conf/sigmod/ArmbrustXLHLBMK15}
\bibfield{author}{\bibinfo{person}{Michael Armbrust}, \bibinfo{person}{Reynold~S. Xin}, \bibinfo{person}{Cheng Lian}, \bibinfo{person}{Yin Huai}, \bibinfo{person}{Davies Liu}, \bibinfo{person}{Joseph~K. Bradley}, \bibinfo{person}{Xiangrui Meng}, \bibinfo{person}{Tomer Kaftan}, \bibinfo{person}{Michael~J. Franklin}, \bibinfo{person}{Ali Ghodsi}, {and} \bibinfo{person}{Matei Zaharia}.} \bibinfo{year}{2015}\natexlab{}.
\newblock \showarticletitle{Spark {SQL:} Relational Data Processing in Spark}. In \bibinfo{booktitle}{\emph{Proceedings of the 2015 {ACM} {SIGMOD} International Conference on Management of Data, Melbourne, Victoria, Australia, May 31 - June 4, 2015}}, \bibfield{editor}{\bibinfo{person}{Timos~K. Sellis}, \bibinfo{person}{Susan~B. Davidson}, {and} \bibinfo{person}{Zachary~G. Ives}} (Eds.). \bibinfo{publisher}{{ACM}}, \bibinfo{pages}{1383--1394}.
\newblock
\urldef\tempurl%
\url{https://doi.org/10.1145/2723372.2742797}
\showDOI{\tempurl}


\bibitem[Behm et~al\mbox{.}(2022)]%
        {DBLP:conf/sigmod/BehmPAACDGHJKLL22}
\bibfield{author}{\bibinfo{person}{Alexander Behm}, \bibinfo{person}{Shoumik Palkar}, \bibinfo{person}{Utkarsh Agarwal}, \bibinfo{person}{Timothy Armstrong}, \bibinfo{person}{David Cashman}, \bibinfo{person}{Ankur Dave}, \bibinfo{person}{Todd Greenstein}, \bibinfo{person}{Shant Hovsepian}, \bibinfo{person}{Ryan Johnson}, \bibinfo{person}{Arvind~Sai Krishnan}, \bibinfo{person}{Paul Leventis}, \bibinfo{person}{Ala Luszczak}, \bibinfo{person}{Prashanth Menon}, \bibinfo{person}{Mostafa Mokhtar}, \bibinfo{person}{Gene Pang}, \bibinfo{person}{Sameer Paranjpye}, \bibinfo{person}{Greg Rahn}, \bibinfo{person}{Bart Samwel}, \bibinfo{person}{Tom van Bussel}, \bibinfo{person}{Herman~Van Hovell}, \bibinfo{person}{Maryann Xue}, \bibinfo{person}{Reynold Xin}, {and} \bibinfo{person}{Matei Zaharia}.} \bibinfo{year}{2022}\natexlab{}.
\newblock \showarticletitle{Photon: {A} Fast Query Engine for Lakehouse Systems}. In \bibinfo{booktitle}{\emph{{SIGMOD} '22: International Conference on Management of Data, Philadelphia, PA, USA, June 12 - 17, 2022}}, \bibfield{editor}{\bibinfo{person}{Zachary~G. Ives}, \bibinfo{person}{Angela Bonifati}, {and} \bibinfo{person}{Amr~El Abbadi}} (Eds.). \bibinfo{publisher}{{ACM}}, \bibinfo{pages}{2326--2339}.
\newblock
\urldef\tempurl%
\url{https://doi.org/10.1145/3514221.3526054}
\showDOI{\tempurl}


\bibitem[Bizer and Schultz(2009)]%
        {DBLP:journals/ijswis/BizerS09}
\bibfield{author}{\bibinfo{person}{Christian Bizer} {and} \bibinfo{person}{Andreas Schultz}.} \bibinfo{year}{2009}\natexlab{}.
\newblock \showarticletitle{The Berlin {SPARQL} Benchmark}.
\newblock \bibinfo{journal}{\emph{Int. J. Semantic Web Inf. Syst.}} \bibinfo{volume}{5}, \bibinfo{number}{2} (\bibinfo{year}{2009}), \bibinfo{pages}{1--24}.
\newblock
\urldef\tempurl%
\url{https://doi.org/10.4018/JSWIS.2009040101}
\showDOI{\tempurl}


\bibitem[Blasgen and Eswaran(1977)]%
        {DBLP:journals/ibmsj/BlasgenE77}
\bibfield{author}{\bibinfo{person}{Mike~W. Blasgen} {and} \bibinfo{person}{Kapali~P. Eswaran}.} \bibinfo{year}{1977}\natexlab{}.
\newblock \showarticletitle{Storage and Access in Relational Data Bases}.
\newblock \bibinfo{journal}{\emph{IBM Systems Journal}} \bibinfo{volume}{16}, \bibinfo{number}{4} (\bibinfo{year}{1977}), \bibinfo{pages}{362--377}.
\newblock


\bibitem[Boncz et~al\mbox{.}(2005)]%
        {DBLP:conf/cidr/BonczZN05}
\bibfield{author}{\bibinfo{person}{Peter~A. Boncz}, \bibinfo{person}{Marcin Zukowski}, {and} \bibinfo{person}{Niels Nes}.} \bibinfo{year}{2005}\natexlab{}.
\newblock \showarticletitle{MonetDB/X100: Hyper-Pipelining Query Execution}. In \bibinfo{booktitle}{\emph{Second Biennial Conference on Innovative Data Systems Research, {CIDR} 2005, Asilomar, CA, USA, January 4-7, 2005, Online Proceedings}}. \bibinfo{publisher}{www.cidrdb.org}, \bibinfo{pages}{225--237}.
\newblock
\urldef\tempurl%
\url{http://cidrdb.org/cidr2005/papers/P19.pdf}
\showURL{%
\tempurl}


\bibitem[Cormode and Muthukrishnan(2005)]%
        {DBLP:journals/jal/CormodeM05}
\bibfield{author}{\bibinfo{person}{Graham Cormode} {and} \bibinfo{person}{S. Muthukrishnan}.} \bibinfo{year}{2005}\natexlab{}.
\newblock \showarticletitle{An improved data stream summary: the count-min sketch and its applications}.
\newblock \bibinfo{journal}{\emph{J. Algorithms}} \bibinfo{volume}{55}, \bibinfo{number}{1} (\bibinfo{year}{2005}), \bibinfo{pages}{58--75}.
\newblock
\urldef\tempurl%
\url{https://doi.org/10.1016/J.JALGOR.2003.12.001}
\showDOI{\tempurl}


\bibitem[Dong et~al\mbox{.}(2017)]%
        {DBLP:conf/cidr/DongCGBSS17}
\bibfield{author}{\bibinfo{person}{Siying Dong}, \bibinfo{person}{Mark Callaghan}, \bibinfo{person}{Leonidas Galanis}, \bibinfo{person}{Dhruba Borthakur}, \bibinfo{person}{Tony Savor}, {and} \bibinfo{person}{Michael Strum}.} \bibinfo{year}{2017}\natexlab{}.
\newblock \showarticletitle{Optimizing Space Amplification in RocksDB}. In \bibinfo{booktitle}{\emph{8th Biennial Conference on Innovative Data Systems Research, {CIDR} 2017, Chaminade, CA, USA, January 8-11, 2017, Online Proceedings}}. \bibinfo{publisher}{www.cidrdb.org}.
\newblock
\urldef\tempurl%
\url{http://cidrdb.org/cidr2017/papers/p82-dong-cidr17.pdf}
\showURL{%
\tempurl}


\bibitem[Graefe(1990)]%
        {10.1145/93605.98720}
\bibfield{author}{\bibinfo{person}{Goetz Graefe}.} \bibinfo{year}{1990}\natexlab{}.
\newblock \showarticletitle{Encapsulation of parallelism in the Volcano query processing system}.
\newblock \bibinfo{journal}{\emph{SIGMOD Rec.}} \bibinfo{volume}{19}, \bibinfo{number}{2} (\bibinfo{date}{May} \bibinfo{year}{1990}), \bibinfo{pages}{102–111}.
\newblock
\showISSN{0163-5808}
\urldef\tempurl%
\url{https://doi.org/10.1145/93605.98720}
\showDOI{\tempurl}


\bibitem[Graefe(1994)]%
        {DBLP:journals/tkde/Graefe94}
\bibfield{author}{\bibinfo{person}{Goetz Graefe}.} \bibinfo{year}{1994}\natexlab{}.
\newblock \showarticletitle{Volcano - An Extensible and Parallel Query Evaluation System}.
\newblock \bibinfo{journal}{\emph{{IEEE} Trans. Knowl. Data Eng.}} \bibinfo{volume}{6}, \bibinfo{number}{1} (\bibinfo{year}{1994}), \bibinfo{pages}{120--135}.
\newblock
\urldef\tempurl%
\url{https://doi.org/10.1109/69.273032}
\showDOI{\tempurl}


\bibitem[Jin et~al\mbox{.}(2023)]%
        {DBLP:conf/cidr/JinFCLS23}
\bibfield{author}{\bibinfo{person}{Guodong Jin}, \bibinfo{person}{Xiyang Feng}, \bibinfo{person}{Ziyi Chen}, \bibinfo{person}{Chang Liu}, {and} \bibinfo{person}{Semih Salihoglu}.} \bibinfo{year}{2023}\natexlab{}.
\newblock \showarticletitle{K{\`{U}}ZU Graph Database Management System}. In \bibinfo{booktitle}{\emph{13th Conference on Innovative Data Systems Research, {CIDR} 2023, Amsterdam, The Netherlands, January 8-11, 2023}}. \bibinfo{publisher}{www.cidrdb.org}.
\newblock
\urldef\tempurl%
\url{https://www.cidrdb.org/cidr2023/papers/p48-jin.pdf}
\showURL{%
\tempurl}


\bibitem[Kersten et~al\mbox{.}(2018)]%
        {DBLP:journals/pvldb/KerstenLKNPB18}
\bibfield{author}{\bibinfo{person}{Timo Kersten}, \bibinfo{person}{Viktor Leis}, \bibinfo{person}{Alfons Kemper}, \bibinfo{person}{Thomas Neumann}, \bibinfo{person}{Andrew Pavlo}, {and} \bibinfo{person}{Peter~A. Boncz}.} \bibinfo{year}{2018}\natexlab{}.
\newblock \showarticletitle{Everything You Always Wanted to Know About Compiled and Vectorized Queries But Were Afraid to Ask}.
\newblock \bibinfo{journal}{\emph{Proc. {VLDB} Endow.}} \bibinfo{volume}{11}, \bibinfo{number}{13} (\bibinfo{year}{2018}), \bibinfo{pages}{2209--2222}.
\newblock
\urldef\tempurl%
\url{https://doi.org/10.14778/3275366.3275370}
\showDOI{\tempurl}


\bibitem[Kikolashvili(2019)]%
        {kikolashvili2019design}
\bibfield{author}{\bibinfo{person}{Giorgi Kikolashvili}.} \bibinfo{year}{2019}\natexlab{}.
\newblock \emph{\bibinfo{title}{On the design of a JVM-based vectorized Spark query engine}}.
\newblock \bibinfo{thesistype}{Ph.\,D. Dissertation}. \bibinfo{school}{Universiteit van Amsterdam}.
\newblock


\bibitem[Lamb et~al\mbox{.}(2012)]%
        {DBLP:journals/pvldb/LambFVTVDB12}
\bibfield{author}{\bibinfo{person}{Andrew Lamb}, \bibinfo{person}{Matt Fuller}, \bibinfo{person}{Ramakrishna Varadarajan}, \bibinfo{person}{Nga Tran}, \bibinfo{person}{Ben Vandiver}, \bibinfo{person}{Lyric Doshi}, {and} \bibinfo{person}{Chuck Bear}.} \bibinfo{year}{2012}\natexlab{}.
\newblock \showarticletitle{The Vertica Analytic Database: C-Store 7 Years Later}.
\newblock \bibinfo{journal}{\emph{Proc. {VLDB} Endow.}} \bibinfo{volume}{5}, \bibinfo{number}{12} (\bibinfo{year}{2012}), \bibinfo{pages}{1790--1801}.
\newblock
\urldef\tempurl%
\url{https://doi.org/10.14778/2367502.2367518}
\showDOI{\tempurl}


\bibitem[Lamb et~al\mbox{.}(2024)]%
        {DBLP:conf/sigmod/LambSHCKHS24}
\bibfield{author}{\bibinfo{person}{Andrew Lamb}, \bibinfo{person}{Yijie Shen}, \bibinfo{person}{Dani{\"{e}}l Heres}, \bibinfo{person}{Jayjeet Chakraborty}, \bibinfo{person}{Mehmet~Ozan Kabak}, \bibinfo{person}{Liang{-}Chi Hsieh}, {and} \bibinfo{person}{Chao Sun}.} \bibinfo{year}{2024}\natexlab{}.
\newblock \showarticletitle{Apache Arrow DataFusion: {A} Fast, Embeddable, Modular Analytic Query Engine}. In \bibinfo{booktitle}{\emph{Companion of the 2024 International Conference on Management of Data, {SIGMOD/PODS} 2024, Santiago AA, Chile, June 9-15, 2024}}, \bibfield{editor}{\bibinfo{person}{Pablo Barcel{\'{o}}}, \bibinfo{person}{Nayat S{\'{a}}nchez{-}Pi}, \bibinfo{person}{Alexandra Meliou}, {and} \bibinfo{person}{S.~Sudarshan}} (Eds.). \bibinfo{publisher}{{ACM}}, \bibinfo{pages}{5--17}.
\newblock
\urldef\tempurl%
\url{https://doi.org/10.1145/3626246.3653368}
\showDOI{\tempurl}


\bibitem[Mhedhbi et~al\mbox{.}(2021)]%
        {DBLP:conf/sigmod/MhedhbiLKWS21}
\bibfield{author}{\bibinfo{person}{Amine Mhedhbi}, \bibinfo{person}{Matteo Lissandrini}, \bibinfo{person}{Laurens Kuiper}, \bibinfo{person}{Jack Waudby}, {and} \bibinfo{person}{G{\'{a}}bor Sz{\'{a}}rnyas}.} \bibinfo{year}{2021}\natexlab{}.
\newblock \showarticletitle{{LSQB:} a large-scale subgraph query benchmark}. In \bibinfo{booktitle}{\emph{{GRADES-NDA} '21: Proceedings of the 4th {ACM} {SIGMOD} Joint International Workshop on Graph Data Management Experiences {\&} Systems {(GRADES)} and Network Data Analytics (NDA), Virtual Event, China, 20 June 2021}}, \bibfield{editor}{\bibinfo{person}{Vasiliki Kalavri} {and} \bibinfo{person}{Nikolay Yakovets}} (Eds.). \bibinfo{publisher}{{ACM}}, \bibinfo{pages}{8:1--8:11}.
\newblock
\urldef\tempurl%
\url{https://doi.org/10.1145/3461837.3464516}
\showDOI{\tempurl}


\bibitem[Neumann(2011)]%
        {DBLP:journals/pvldb/Neumann11}
\bibfield{author}{\bibinfo{person}{Thomas Neumann}.} \bibinfo{year}{2011}\natexlab{}.
\newblock \showarticletitle{Efficiently Compiling Efficient Query Plans for Modern Hardware}.
\newblock \bibinfo{journal}{\emph{Proc. {VLDB} Endow.}} \bibinfo{volume}{4}, \bibinfo{number}{9} (\bibinfo{year}{2011}), \bibinfo{pages}{539--550}.
\newblock
\urldef\tempurl%
\url{https://doi.org/10.14778/2002938.2002940}
\showDOI{\tempurl}


\bibitem[Neumann and Moerkotte(2011)]%
        {DBLP:conf/icde/NeumannM11}
\bibfield{author}{\bibinfo{person}{Thomas Neumann} {and} \bibinfo{person}{Guido Moerkotte}.} \bibinfo{year}{2011}\natexlab{}.
\newblock \showarticletitle{Characteristic sets: Accurate cardinality estimation for {RDF} queries with multiple joins}. In \bibinfo{booktitle}{\emph{Proceedings of the 27th International Conference on Data Engineering, {ICDE} 2011, April 11-16, 2011, Hannover, Germany}}, \bibfield{editor}{\bibinfo{person}{Serge Abiteboul}, \bibinfo{person}{Klemens B{\"{o}}hm}, \bibinfo{person}{Christoph Koch}, {and} \bibinfo{person}{Kian{-}Lee Tan}} (Eds.). \bibinfo{publisher}{{IEEE} Computer Society}, \bibinfo{pages}{984--994}.
\newblock
\urldef\tempurl%
\url{https://doi.org/10.1109/ICDE.2011.5767868}
\showDOI{\tempurl}


\bibitem[Neumann and Weikum(2008)]%
        {DBLP:journals/pvldb/NeumannW08}
\bibfield{author}{\bibinfo{person}{Thomas Neumann} {and} \bibinfo{person}{Gerhard Weikum}.} \bibinfo{year}{2008}\natexlab{}.
\newblock \showarticletitle{{RDF-3X:} a RISC-style engine for {RDF}}.
\newblock \bibinfo{journal}{\emph{Proc. {VLDB} Endow.}} \bibinfo{volume}{1}, \bibinfo{number}{1} (\bibinfo{year}{2008}), \bibinfo{pages}{647--659}.
\newblock
\urldef\tempurl%
\url{https://doi.org/10.14778/1453856.1453927}
\showDOI{\tempurl}


\bibitem[Ngom et~al\mbox{.}(2021)]%
        {DBLP:conf/damon/NgomMB0LMP21}
\bibfield{author}{\bibinfo{person}{Amadou Ngom}, \bibinfo{person}{Prashanth Menon}, \bibinfo{person}{Matthew Butrovich}, \bibinfo{person}{Lin Ma}, \bibinfo{person}{Wan~Shen Lim}, \bibinfo{person}{Todd~C. Mowry}, {and} \bibinfo{person}{Andrew Pavlo}.} \bibinfo{year}{2021}\natexlab{}.
\newblock \showarticletitle{Filter Representation in Vectorized Query Execution}. In \bibinfo{booktitle}{\emph{Proceedings of the 17th International Workshop on Data Management on New Hardware, DaMoN 2021, 21 June 2021, Virtual Event, China}}, \bibfield{editor}{\bibinfo{person}{Danica Porobic} {and} \bibinfo{person}{Spyros Blanas}} (Eds.). \bibinfo{publisher}{{ACM}}, \bibinfo{pages}{6:1--6:7}.
\newblock
\urldef\tempurl%
\url{https://doi.org/10.1145/3465998.3466009}
\showDOI{\tempurl}


\bibitem[Pedreira et~al\mbox{.}(2022)]%
        {DBLP:journals/pvldb/PedreiraEBWSPHC22}
\bibfield{author}{\bibinfo{person}{Pedro Pedreira}, \bibinfo{person}{Orri Erling}, \bibinfo{person}{Maria Basmanova}, \bibinfo{person}{Kevin Wilfong}, \bibinfo{person}{Laith Sakka}, \bibinfo{person}{Krishna Pai}, \bibinfo{person}{Wei He}, {and} \bibinfo{person}{Biswapesh Chattopadhyay}.} \bibinfo{year}{2022}\natexlab{}.
\newblock \showarticletitle{Velox: Meta's Unified Execution Engine}.
\newblock \bibinfo{journal}{\emph{Proc. {VLDB} Endow.}} \bibinfo{volume}{15}, \bibinfo{number}{12} (\bibinfo{year}{2022}), \bibinfo{pages}{3372--3384}.
\newblock
\urldef\tempurl%
\url{https://doi.org/10.14778/3554821.3554829}
\showDOI{\tempurl}


\bibitem[Raasveldt(2022)]%
        {DBLP:conf/vldb/Raasveldt22}
\bibfield{author}{\bibinfo{person}{Mark Raasveldt}.} \bibinfo{year}{2022}\natexlab{}.
\newblock \showarticletitle{DuckDB - {A} Modern Modular and Extensible Database System}. In \bibinfo{booktitle}{\emph{1st International Workshop on Composable Data Management Systems, CDMS@VLDB 2022, Sydney, Australia, September 9, 2022}}, \bibfield{editor}{\bibinfo{person}{Satyanarayana~R. Valluri} {and} \bibinfo{person}{Mohamed Za{\"{\i}}t}} (Eds.).
\newblock
\urldef\tempurl%
\url{https://cdmsworkshop.github.io/2022/Proceedings/Keynotes/Abstract\_MarkRaasveldt.pdf}
\showURL{%
\tempurl}


\bibitem[Taft et~al\mbox{.}(2020)]%
        {DBLP:conf/sigmod/TaftSMVLGNWBPBR20}
\bibfield{author}{\bibinfo{person}{Rebecca Taft}, \bibinfo{person}{Irfan Sharif}, \bibinfo{person}{Andrei Matei}, \bibinfo{person}{Nathan VanBenschoten}, \bibinfo{person}{Jordan Lewis}, \bibinfo{person}{Tobias Grieger}, \bibinfo{person}{Kai Niemi}, \bibinfo{person}{Andy Woods}, \bibinfo{person}{Anne Birzin}, \bibinfo{person}{Raphael Poss}, \bibinfo{person}{Paul Bardea}, \bibinfo{person}{Amruta Ranade}, \bibinfo{person}{Ben Darnell}, \bibinfo{person}{Bram Gruneir}, \bibinfo{person}{Justin Jaffray}, \bibinfo{person}{Lucy Zhang}, {and} \bibinfo{person}{Peter Mattis}.} \bibinfo{year}{2020}\natexlab{}.
\newblock \showarticletitle{CockroachDB: The Resilient Geo-Distributed {SQL} Database}. In \bibinfo{booktitle}{\emph{Proceedings of the 2020 International Conference on Management of Data, {SIGMOD} Conference 2020, online conference [Portland, OR, USA], June 14-19, 2020}}, \bibfield{editor}{\bibinfo{person}{David Maier}, \bibinfo{person}{Rachel Pottinger}, \bibinfo{person}{AnHai Doan}, \bibinfo{person}{Wang{-}Chiew Tan}, \bibinfo{person}{Abdussalam Alawini}, {and} \bibinfo{person}{Hung~Q. Ngo}} (Eds.). \bibinfo{publisher}{{ACM}}, \bibinfo{pages}{1493--1509}.
\newblock
\urldef\tempurl%
\url{https://doi.org/10.1145/3318464.3386134}
\showDOI{\tempurl}


\bibitem[Utsin(2019)]%
        {CockroachDBVectorizedMJ}
\bibfield{author}{\bibinfo{person}{George Utsin}.} \bibinfo{year}{2019}\natexlab{}.
\newblock \bibinfo{booktitle}{\emph{Vectorizing the merge joiner in CockroachDB}}.
\newblock
\urldef\tempurl%
\url{https://www.cockroachlabs.com/blog/vectorizing-the-merge-joiner-in-cockroachdb/}
\showURL{%
Retrieved November 26, 2024 from \tempurl}


\bibitem[Zukowski et~al\mbox{.}(2008)]%
        {DBLP:conf/damon/ZukowskiNB08}
\bibfield{author}{\bibinfo{person}{Marcin Zukowski}, \bibinfo{person}{Niels Nes}, {and} \bibinfo{person}{Peter~A. Boncz}.} \bibinfo{year}{2008}\natexlab{}.
\newblock \showarticletitle{{DSM} vs. {NSM:} {CPU} performance tradeoffs in block-oriented query processing}. In \bibinfo{booktitle}{\emph{4th Workshop on Data Management on New Hardware, DaMoN 2008, Vancouver, BC, Canada, June 13, 2008}}, \bibfield{editor}{\bibinfo{person}{Qiong Luo} {and} \bibinfo{person}{Kenneth~A. Ross}} (Eds.). \bibinfo{publisher}{{ACM}}, \bibinfo{pages}{47--54}.
\newblock
\urldef\tempurl%
\url{https://doi.org/10.1145/1457150.1457160}
\showDOI{\tempurl}


\end{thebibliography}

\end{document}